\documentclass[twocolumn,12pt]{aastex6}

\newcommand{\chandra}{{\it Chandra}}

\slugcomment{To appear in the {\em Astrophysical Journal}}
\lefthead{Drake et al.}
\righthead{\chandra\ spectroscopy of V745 Scorpii}

\begin{document}

\title{Collimation and asymmetry of the hot blast wave from the recurrent nova V745 Sco}

\author{Jeremy J. Drake,\altaffilmark{1}
Laura Delgado,\altaffilmark{2}
J. Martin Laming,\altaffilmark{3}
Sumner Starrfield,\altaffilmark{4}
Vinay~Kashyap,\altaffilmark{1}
Salvatore Orlando,\altaffilmark{5}
Kim L. Page,\altaffilmark{6}
M.~Hernanz,\altaffilmark{2}
J-U.~Ness,\altaffilmark{7}
R.~D.~Gehrz,\altaffilmark{8}
Daan van Rossum\altaffilmark{9}
and Charles E.~Woodward\altaffilmark{8}}

\affil{$^1$Harvard-Smithsonian Center for Astrophysics, 
60 Garden Street, Cambridge, MA 02138\\
$^{2}$Institute of Space Sciences, ICE (CSIC-IEEC), 08193 Cerdanyola del Vallés, Barcelona, Spain\\
$^3$Space Science Division,
Naval Research Laboratory, Code 7674L, Washington DC 20375\\
$^4$School of Earth and Space Exploration, Arizona
State University, Tempe, AZ 85287-1404, USA\\
$^5$INAF - Osservatorio Astronomico di Palermo G. S. Vaiana, Piazza del Parlamento 1, 90134 Palermo, Italy\\
$^6$Department of Physics \& Astronomy, University of
Leicester, Leicester, LE1 7RH, UK\\
$^{7}$Science Operations Division, Science Operations Department of ESA, ESAC, 28691 Villanueva de la Ca\~nada (Madrid), Spain\\
$^{8}$Minnesota Institute for Astrophysics, School of Physics and Astronomy, University of Minnesota, 116 Church Street S.E., Minneapolis, MN 55455\\
$^{9}$Flash Center for Computational Science, Department of Astronomy and Astrophysics, University of Chicago, Chicago, IL 60637, USA}

\begin{abstract}
The recurrent symbiotic nova V745~Sco exploded on 2014 February 6 and
was observed on February 22 and 23 by the {\it Chandra} X-ray
Observatory Transmission Grating Spectrometers.  
By that time the supersoft source phase had
already ended and {\it Chandra} spectra are consistent with emission
from a hot, shock-heated circumstellar medium with temperatures
exceeding $10^7$~K.  X-ray line profiles are more sharply peaked
than expected for a spherically-symmetric blast wave, with a full
width at zero intensity of approximately 2400~km~s$^{-1}$, a full
width at half maximum of $1200\pm 30$~km~s$^{-1}$ and an average net
blueshift of $165 \pm 10$~km~s$^{-1}$.  The red wings of lines are
increasingly absorbed  toward longer wavelengths by material within the remnant.
We conclude that the blast
wave was sculpted by an aspherical circumstellar medium in which an equatorial
density enhancement plays a role, as in earlier symbiotic
nova explosions.  Expansion of the dominant X-ray emitting material is
aligned close to the plane of the sky and most
consistent with an orbit seen close to face-on.  Comparison of an
analytical blast wave model with the X-ray spectra, {\it Swift} observations  and near-infrared line widths indicates the explosion
energy was approximately $10^{43}$~erg, and confirms an ejected mass
of approximately $10^{-7}M_\odot$.  The total mass lost is an order of magnitude lower than the accreted mass required to have initiated the explosion, indicating the white dwarf is gaining mass and is a supernova Type~1a progenitor candidate. 
\end{abstract}


\section{INTRODUCTION}
\label{s:intro}

V745 Sco is a member of  the exclusive class of cataclysmic variables known as recurrent symbiotic novae. Symbiotic novae are close binaries in which a white dwarf orbits within the wind or extended atmosphere of an evolved companion.   Accretion onto the white dwarf from the ambient wind, or through Roche Lobe overflow from a disk, leads to a build-up of matter that reaches sufficient temperature and density to initiate a thermonuclear runaway \citep[TNR;][]{Starrfield.etal:74}.  Recurrent novae (RNe) are cases of generally more massive white dwarfs with accretion rates approaching that required for steady surface nuclear burning \citep[see, e.g.,][]{Sugimoto.Miyaji:81,Starrfield.etal:88,Nomoto.etal:07,Shen.Bildsten:07,Wolf.etal:13} that engender outbursts at quasi-regular intervals of only a few years.  There are only 10 RNe presently known in the Milky Way and most have evolved binary companions \citep[see][]{Schaefer:10}. 

Nova explosions in symbiotic systems such as V745~Sco have been likened to mini versions of Type~II supernovae \citep{Bode.Kahn:85}.  A blast wave propagates through the ambient circumstellar material and red giant wind in much the same way as a core-collapse supernova blast propagates through the wind of its massive progenitor.   The main difference lies in the explosion energy and ejected mass: recurrent novae (a sub-class of classical novae) are typically about $10^5$ times less energetic and expel $10^5$--$10^7$ times less mass. 
Consequently, the blast evolves on much shorter timescales than a supernova---weeks instead of millennia \citep{Bode.Kahn:85}---and this evolution can, at least in principle, be observed in considerable detail.  The scarcity of known RNe renders each outburst a valuable opportunity to study the progenitor system, the nature of the explosion and its complex interaction with its circumstellar environment. 

A considerable body of evidence now points to significant structure and asymmetry in nova blast waves.  \citet{Hutchings:72} discussed spectroscopic line profiles for three different nova events and deduced deviations from spherical symmetry in the ejecta distributions. 
VLA radio observations of the 1985 outburst of RS~Oph by \citet{Hjellming.etal:86} provided some hints of deviation from sphericity that were later confirmed with VLBI observations by \citet{Taylor.etal:89}.   A bipolar morphology was also indicated by both radio and optical imaging of the 2006 RS~Oph outburst \citep{O'Brien.etal:06,Bode.etal:07,Sokoloski.etal:08,Ribeiro.etal:09}, while \citet{Drake.etal:09} found the signature of a collimated blast in high-resolution {\it Chandra} X-ray spectra that provided a direct probe of the shock-heated plasma.  Detailed multi-dimensional hydrodynamic simulations have confirmed that collimation and asymmetry arise from interaction of the explosion with circumstellar material---either an accretion disk, a companion wind, or both \citep{Walder.etal:08,Orlando.etal:09,Drake.Orlando:10,Orlando.Drake:12,Pan.etal:15}.  \citet{Ness.etal:13} also found evidence that high-inclination novae in the super-soft source (SSS) phase exhibit more soft X-ray emission lines than low-inclination systems; the lines are likely due to reprocessing by obscuration and aspherical ejecta.

The {\it Chandra} high-resolution X-ray spectrometers were deployed to observe the 2014 V745~Sco explosion and investigate the nature of the blast wave and super-soft source.  As we relate below, the SSS faded so quickly that it had disappeared by the time our observations could be made.  Exquisite X-ray spectra of the blast wave were obtained, however.  Here, we analyze the unique constraints on the blast wave conditions and geometry afforded by the {\it Chandra} data, complemented by monitoring observations made by {\it Swift} and near-infrared Paschen line widths culled from the literature. 

\section{V745 Sco}
\label{s:v745}

V745~Sco was discovered in outburst at visual magnitude 9 by 
Rod Stubbings at Tetoora Road Observatory, Victoria, Australia on 2014 February 6.694 UT \citep{Waagen:14}.  Only two previous outbursts, in 1937 and 1989, had been recorded for this nova \citep[e.g.,][]{Duerbeck:89}, although the 50 year inter-outburst interval compared with the 25 year interval separating the 1989 and 2014 outbursts support the conjecture of \citet{Schaefer:10} that an outburst was missed in the 1960s and that another would be due close to 2013.   It is a very ``fast" nova, evolving and fading by two and three visual magnitudes in only 6.2 and 9 days, respectively \citep{Banerjee.etal:14}.  

The explosion triggered observing campaigns over the entire electromagnetic spectrum, including radio \citep{Rupen.etal:14,Kantharia.etal:15}, infrared \citep{Banerjee.etal:14}, optical  \citep{Anupama.etal:14,Mroz.etal:14}
X-ray \citep{Mukai.etal:14,Beardmore.etal:14,Page.etal:15,Drake.etal:14}, hard X-ray  \citep{Rana.etal:14,Orio.etal:15} and $\gamma$-ray \citep{Cheung.etal:14,Cheung.etal:15} wavelengths.  \citet{Page.etal:15} present a detailed summary of the various observations.  

There are several noteworthy aspects of the event from the perspective of the work presented here.  Foremost are the {\it Fermi} LAT $2\sigma$ and $3\sigma$ $\gamma$-ray detections on 2014 February 6 and 7 (days 1 and 2 of the outburst) by \citet{Cheung.etal:14}---the review of \citet{Cheung.etal:15} notes that V745~Sco is only the sixth nova to have been detected in $\gamma$-rays.
The rise on day 3 and fall on day 10 of the SSS phase \citep{Page.etal:14,Page.etal:15} was remarkably fast and among the shortest ever observed.
{\em NuSTAR} observations detected X-ray emission up to 20~keV and  
indicated the presence of a hot plasma with an average temperature of 2.7~keV  \citep{Orio.etal:15}.   Greatly broadened optical and infrared line profiles indicating expansion velocities of more than 4000~km~s$^{-1}$ were observed by \citet{Anupama.etal:14} and \citet{Banerjee.etal:14}, reminiscent of the 1989 outburst \citep{Duerbeck:89,Williams.etal:03,Wagner.Starrfield:14}. 

Relatively little is known about the V745~Sco stellar system because it lies toward the crowded Galactic bulge region.  \citet{Duerbeck:89} classified the red giant companion as M6~III based on TiO bands, while \citet{Harrison.etal:93} inferred a spectral type of M4~III based on CO absorption features.  \citet{Schaefer:10} derived an orbital period of $510 \pm 20$~days using optical photometry, and a distance of $7.8 \pm1.2$~kpc.  The period has been refuted by \citet{Mroz.etal:14}, who failed to find evidence for a binary period in extensive Optical Gravitational Lensing Experiment (OGLE) data.  They note that the 
quiescence variability of the red giant is characterized by semi-regular pulsations with periods of 136.5 and 77.4 days.

\section{OBSERVATIONS}
\label{s:obs}

\begin{figure*}
\begin{center}
\includegraphics[angle=0,width=6.in]{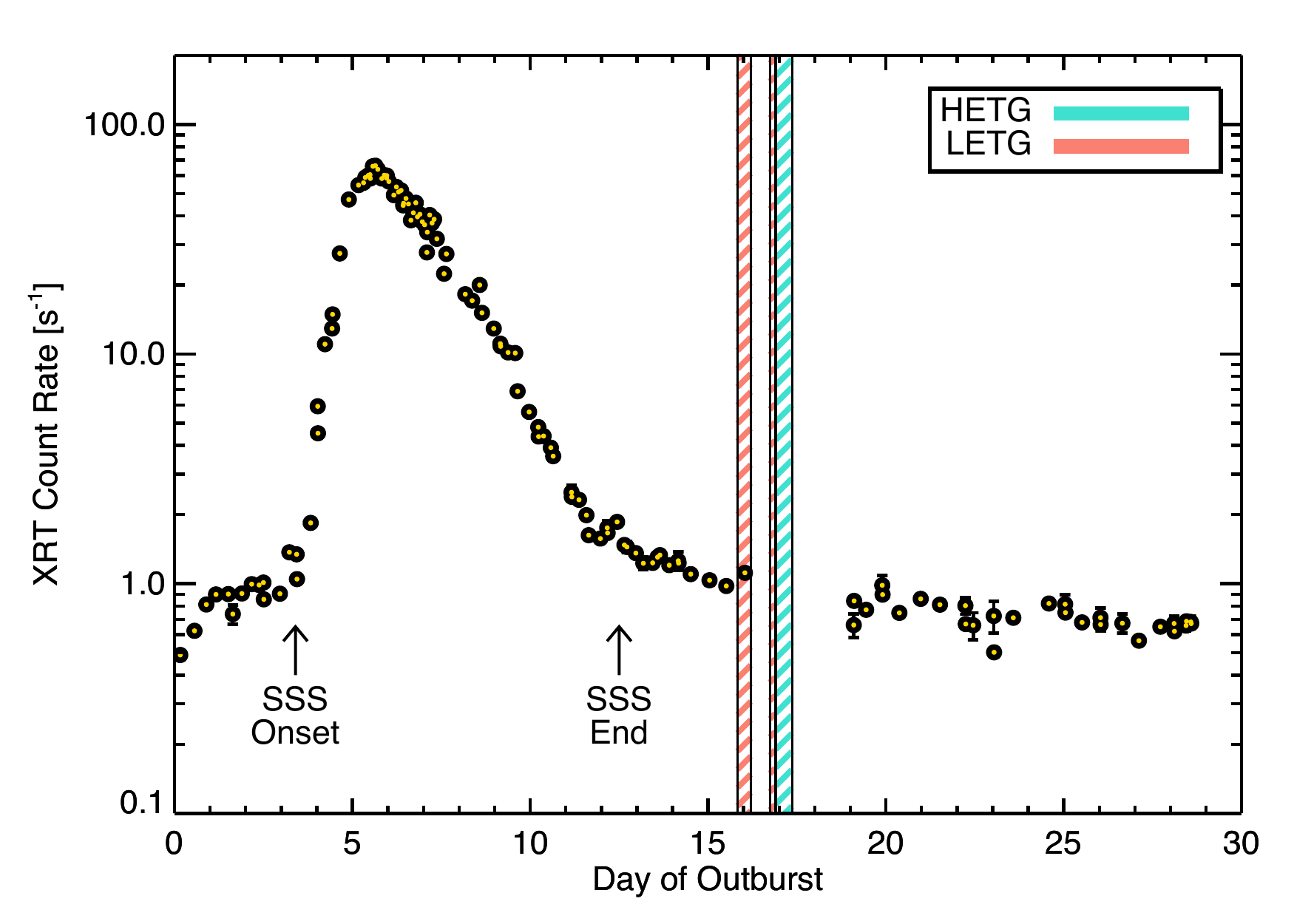}
\end{center}
\caption{The {\it Swift} XRT count rate for V745~Sco as a function of the time since the outburst was first identified.  The times of {\it Chandra} high-resolution spectroscopic observations are indicated by the shaded regions. The times of SSS onset and end discussed by \citet{Page.etal:15} are also illustrated.
\label{f:swift}}
\end{figure*}

V745 Sco was observed by the {\it Chandra} X-ray Observatory \citep{Weisskopf.etal:03} Low Energy (LETG; \citealt{Brinkman.etal:00})
and High Energy (HETG; \citealt{Canizares.etal:00}) transmission grating spectrometers on UT 2014 February 22, and 23, with net exposure
times of 45 and 39~ks, respectively. The LETG observations employed the High Resolution Camera Spectroscopy array (HRC-S) and comprised two segments of approximately 32~ks and 13~ks with start times separated by about 20 hours.  The HETG observation, using the Advanced CCD Imaging Spectrometer spectroscopic detector (ACIS-S), followed on the heels of the latter.  {\it Chandra} observations are illustrated in the context of the evolving X-ray light obtained by the {\it Swift} observatory \citep{Page.etal:15} in Figure~\ref{f:swift}.
Details of the {\it Chandra} observations are reported in Table~\ref{t:obs}.  

\begin{table*}[htdp]
\caption{Details of {\it Chandra} high-resolution spectroscopic observations of V745 Sco}
\label{t:obs}
\begin{center}
\begin{tabular}{lccccr}
\hline
ObsID & Instrument & Day\tablenotemark{1} & $t_{\rm start}$ & $t_{\rm stop}$ & Exp.~(s) \\ \hline     \hline
15738 & LETG+HRC-S  & 15.8 & 2014-02-22 12:30:19 & 2014-02-22 21:49:01 & 32236 \\
16595 & LETG+HRC-S  & 16.7 & 2014-02-23 10:20:13 & 2014-02-23 14:17:32 & 12937 \\
15737 & HETG+ACIS-S & 16.9 & 2014-02-23 14:17:32 & 2014-02-24 02:02:51 & 39457 \\
\hline
\end{tabular}
\tablenotetext{1}{Day of outburst at observation start, based on an assumed explosion initiation on the discovery date of 2014-2-06.694 UT (JD 2456695.194).}
\end{center}
\label{default}
\end{table*}%

\begin{figure*}
\begin{center}
\includegraphics[angle=0,width=7.in]{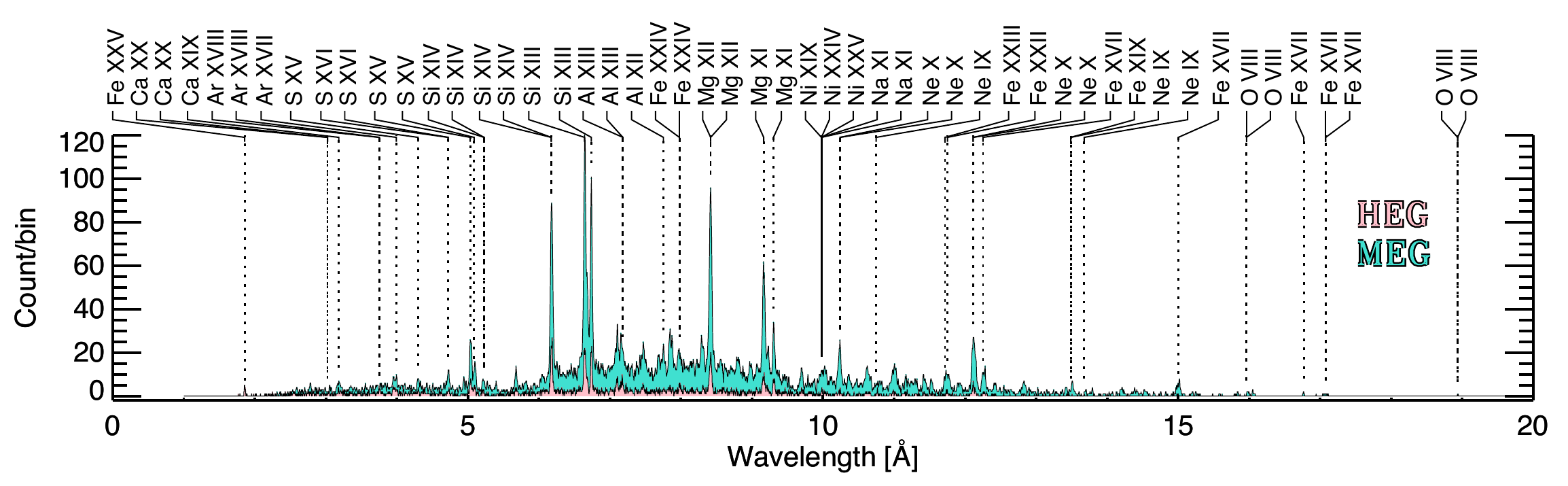}
\includegraphics[angle=0,width=7.in]{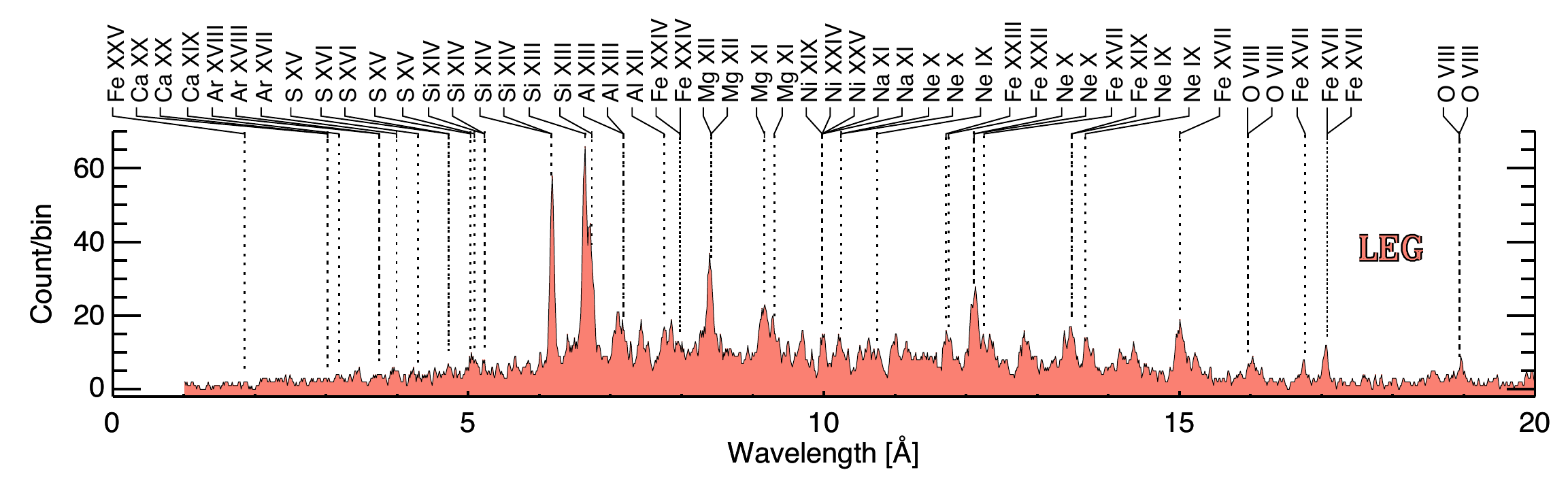}
\end{center}
\caption{{\it Chandra} High Energy Grating (HEG) Medium Energy Grating (MEG; top) and Low Energy Grating (LEG; bottom) spectra of V745 Sco obtained between 15.8 and 17.4 days into the outburst.  Data are shown as the total accumulated counts per spectral bin, where the bin sizes are 0.0025~\AA\ and 0.005~\AA\ for HEG and MEG, respectively, and 0.0125~\AA\ for LEG. Identifications of prominent spectral lines are indicated.
\label{f:lineid}}
\end{figure*}

Data were processed using standard procedures\footnote{http://cxc.harvard.edu/ciao/threads} and {\it Chandra} calibration database version 4.6.8.   The extracted HETG spectra with prominent spectral lines identified are illustrated in Figure~\ref{f:lineid}.

{\it Swift} initiated observations of V745~Sco 3.7~hr after the optical discovery, using the X- ray Telescope (XRT; \citealt{Burrows.etal:05}) and UV/Optical Telescope (UVOT; \citealt{Roming.etal:05}).  Data were obtained several times a day on most of the first 28~days of the outburst, and regular observations were continued at a lower cadence until the end of 2014 September.  A full description of the {\it Swift} campaign, data processing and analysis has been presented by \citet{Page.etal:15}.  We concentrate here on the first 40 days of observations of the blast wave emission; the reader is referred to  \citet{Page.etal:15} for an analysis of the SSS phase.  

\section{ANALYSIS}
\label{s:anal}

Analysis of the {\it Chandra} spectra involved two separate approaches: a model parameter estimation method to examine the condition of the shocked material in the blast wave, and an analysis of spectral line profiles in order to investigate possible constraints on the nature and geometry of the explosion.

\subsection{Characterizing the shocked gas using model parameter estimation} 
\label{s:fitting}
 
\subsubsection{General approach}

Parameter estimation was performed using the {\it Sherpa} fitting engine within the {\it Chandra} Interactive Analysis of Observations (CIAO; \citealt{Fruscione.etal:06}) software framework version 4.7.  The {\it Chandra} spectra in Figure~\ref{f:lineid} show bright emission lines of cosmically-abundant elements on top of a broad continuum and bear a remarkable resemblance to those obtained of the 2006 RS~Oph explosion \citep{Drake.etal:09,Ness.etal:09}.  Indeed, based on the available information noted in Section~\ref{s:v745}, V745~Sco is likely to be a very similar system to RS~Oph.  Blast wave X-ray spectra of symbiotic novae are expected to originate from the diffuse shock-heated secondary star wind, and the RS~Oph blast wave proved to be well-described by optically-thin, collision-dominated plasma emission \citep{Bode.etal:06,Nelson.etal:08,Drake.etal:09,Ness.etal:09}. We therefore proceeded to model the V754~Sco spectra using thermal plasma radiative loss models.  As a baseline, Astrophysical Plasma Emission Code (APEC\footnote{http://www.atomdb.org/}) 
models were adopted with their default (and traditional) ``solar'' abundances of  \citet{Anders.Grevesse:89}.  

Since high-resolution {\it Chandra} HETG spectra, those of V745~Sco being no exception, often comprise many bins with few counts, minimization of model deviations from the data employed the Cash statistic (\citealt{Cash:79}; we denote the reduced form here by $\mathcal C_r$), which is valid in the Poisson regime of low numbers of counts in which $\chi^2$ approaches, relying on Gaussian uncertainties, are inapplicable.   This allows the data to be analyzed without further grouping of neighboring bins, and at full spectral resolution.  The HETG High Energy Grating (HEG) spectrum was fit over the range 1.5--16~\AA, while fits to the Medium Energy Grating (MEG) spectrum were restricted to 2--20~\AA. Both HEG and MEG spectra were fit simultaneously.  

The deployment of the LETG+HRC-S sought to capture the SSS emission, which was expected to dominate the X-ray signal longward of 20~\AA\ or so.  The startling rapidity with which V745~Sco evolved meant that the SSS had already faded by the time the deployment was made.  Nevertheless, the LETG data are still capable of providing an independent measure of the blast wave conditions, albeit with lower precision than the HETG observations.

LETG+HRC-S spectra are subject to higher levels of background than the HETG+ACIS-S combination, largely owing to the anti-coincidence shield being non-functional on-orbit.   Parameter estimation for LETGS spectra fitting both source and background simultaneously encountered software problems.  Consequently, we 
employed background subtraction and binning of the data to a minimum of 10 counts per bin, combined with the \citet{Gehrels:86} modification to the $\chi^2$ statistic as a goodness of fit measure.  The higher background level and lower quantum efficiency of the HRC-S compared with ACIS-S, combined with the lower spectral resolution of the LETG in the wavelength range of interest, meant that  the LETG data provided less stringent constraints on model parameters. 

\citet{Banerjee.etal:14} discussed the evolution of the width of the Pa$\beta$ line over the first two weeks of the outburst, showing that on day 15 the full-width at half-maximum intensity (FWHM) still exceeded 1000~km~s$^{-1}$.  The resolving power (FWHM) of the HEG is about 1000 at 12~\AA, corresponding to 300~km~s$^{-1}$.  The blast wave HETG spectra are therefore likely to be significantly broadened, as were those of the recurrent nova RS~Oph \citep{Drake.etal:09}.  Velocity broadening, as well as a net ``redshift'', were therefore also included as free parameters. 

Representative parameter estimation results together with $1\sigma$ (68.3\%) confidence bounds are listed for HETG and LETG data in Tables~\ref{t:hfit_results} and \ref{t:lfit_results}, respectively.  Uncertainties are based only on the statistical properties of the model and data comparisons and do not include errors resulting from instrument calibration or spectral model input data.
 
In order of increasing complexity, data were first matched to an absorbed isothermal model and later to more complex models that included multiple temperature plasma components
and variable element abundances.  In the absorbed isothermal model, the abundances of cosmically-abundant elements were allowed to vary together as a global metallicity.   Absorption was first treated as a simple, single interstellar medium component represented by the column density of neutral hydrogen proxy, $N_H$, and H and He ionization fractions of 0.1 and 0.01, respectively.   This provides a gross and simple characterization of the bulk of the emission, although the fit was relatively poor in terms of reduced fit statistic ($\mathcal C_r=1.38$).  
Subsequent models with multiple temperature plasma components and variable element
abundances also included a more realistic treatment of absorption.

\subsubsection{Absorption model}

Spectral line profiles betray the presence of significant self-absorption: red wings are weakened with respect to their blue counterparts (see Section~\ref{s:profiles} below for further discussion). Such self-absorption was first seen in the {\it Chandra} HETG spectrum of the RS~Ophiuchi blast wave \citep{Drake.etal:09} and arises because the red wings correspond to emission from receding plasma, located on the far side of the blast.  This emission is absorbed by material within the blast wave, in addition to the interstellar medium.  A single representative absorbing column density results in too much attenuation at longer wavelengths that obliterates model predictions of the weak, but still present, hydrogen-like O emission near 19~\AA, even in more complex multi-thermal models (described below), and similarly renders predictions of transitions such as the prominent Fe~XVII 15~\AA\ line much weaker than observed.  A more complete and accurate description of the absorption of the HETG spectrum would involve integration over the material in the line of sight to all parts of both the approaching and receding shocked plasma.  

Detailed hydrodynamic simulations of the RS~Oph blast \citep{Orlando.etal:09,Walder.etal:08} revealed a very complex shock and ejecta structure collimated and deflected by the accretion disk and other circumstellar material, and a highly asymmetric distribution of X-ray emission.  In the absence of detailed information on the V745~Sco blast wave morphology, there are few constraints on a compound absorption model.  We therefore adopted a simple covering fraction varying as a power-law distribution of the column density following \citet{Norton.etal:91} and \citet{Done.Magdziarz:98}, as implemented in the {\it pwab} model within {\it Sherpa}.   Such an absorption model has an effective transmittance as a function of energy, $T(E)$, given by 
\begin{equation}
T(E)=a \int_{N_{Hmin}}^{N_{Hmax}} N_H^\beta exp[-N_H \sigma(E)]\, dN_H,
\end{equation}
where $\sigma(E)$ is the absorption cross-section as a function of energy and $a$ is a constant depending on the power-law index $\beta$ and the minimum and maximum column densities considered ($N_{Hmin}$ and ${N_{Hmax}}$), such that the total covering fraction as a function of column density, $C_f(N_H)\propto N_H^\beta$, is normalized to 1 (see \citealt{Done.Magdziarz:98} Eqn.~1).   

Test fits revealed little sensitivity in terms of model fit quality to the exact minimum and maximum absorbing columns.  Model fits using a single absorbing column retrieved best-fit neutral hydrogen column densities in the general range $N_H\sim 4$--$10\times 10^{21}$cm$^{-2}$.  This is slightly higher than expected from the trend $N_H\propto t^{-0.76\pm0.10}$ by \citet{Page.etal:15} based on observations over the first three days, suggesting that the decrease in absorption had leveled off before day 16.
 Values  of the limiting columns in the complex absorber 
were based on an isothermal model fit to the data in which the $N_H$ limits were allowed to vary.  This trial  resulted in lower and upper $1\sigma$ confidence limits of $N_{Hmin}\geq 2 \times 10^{21}$ and $N_{Hmax}\leq 1.9\times 10^{22}$~cm$^{-2}$. 
These also straddle the range of values found from single absorber fits, and fixed values of $N_{Hmin}=2 \times 10^{21}$ and $N_{Hmax}=2\times 10^{22}$~cm$^{-2}$ were adopted in subsequent model parameter estimation.
 
\subsubsection{Chemical composition}
\label{s:abuns}

The most prominent spectral lines in the HEG, MEG and short wavelength end of the LEG spectra originate from the H-like and He-like ions of the abundant metals O, Ne, Mg, Al, Si, S, Ar and Fe.  In order to investigate possible deviations from a solar abundance mixture, isothermal, two-temperature and three-temperature fits were performed in which the relative abundances of these elements were  allowed to vary independently.  Exceptions were the abundances of Ca and Ni, which were tied to that of Fe, while the S abundance was tied to that of Ar.  These abundance ties reduced the number of free model parameters for quantities that do not play an important role in the data. 
An isothermal model with these constraints fitted to the HEG and MEG spectra had a reduced statistic of $\mathcal C_r=1.30$, while a similar model fit to LEG data had $\mathcal C_r=0.84$.  

\begin{figure*}
\begin{center}
\includegraphics[width=6.5in]{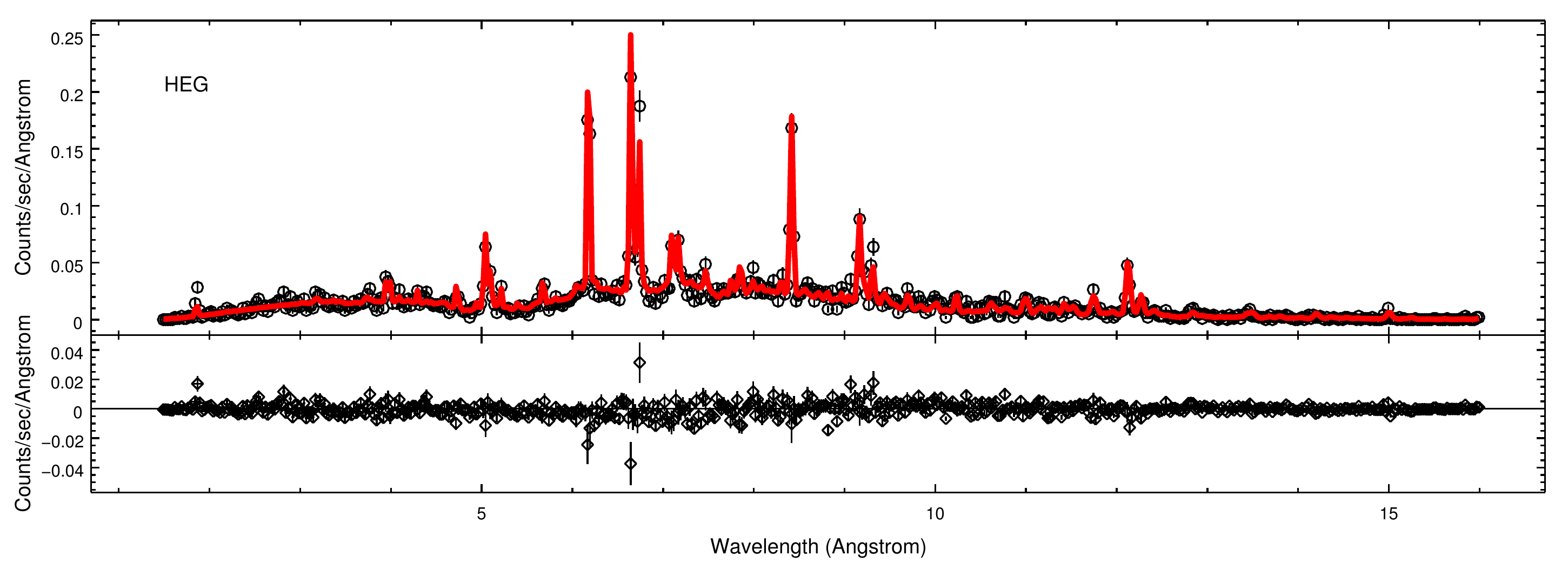}
\includegraphics[width=6.5in]{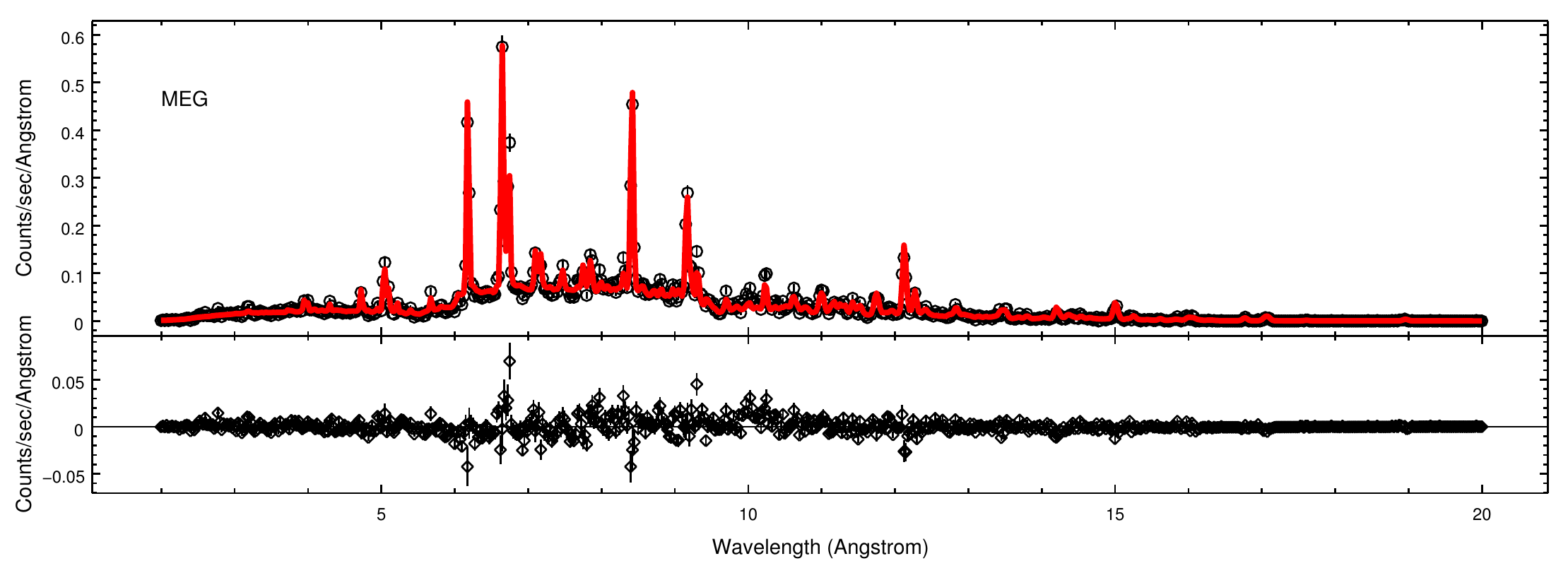}
\includegraphics[width=6.5in]{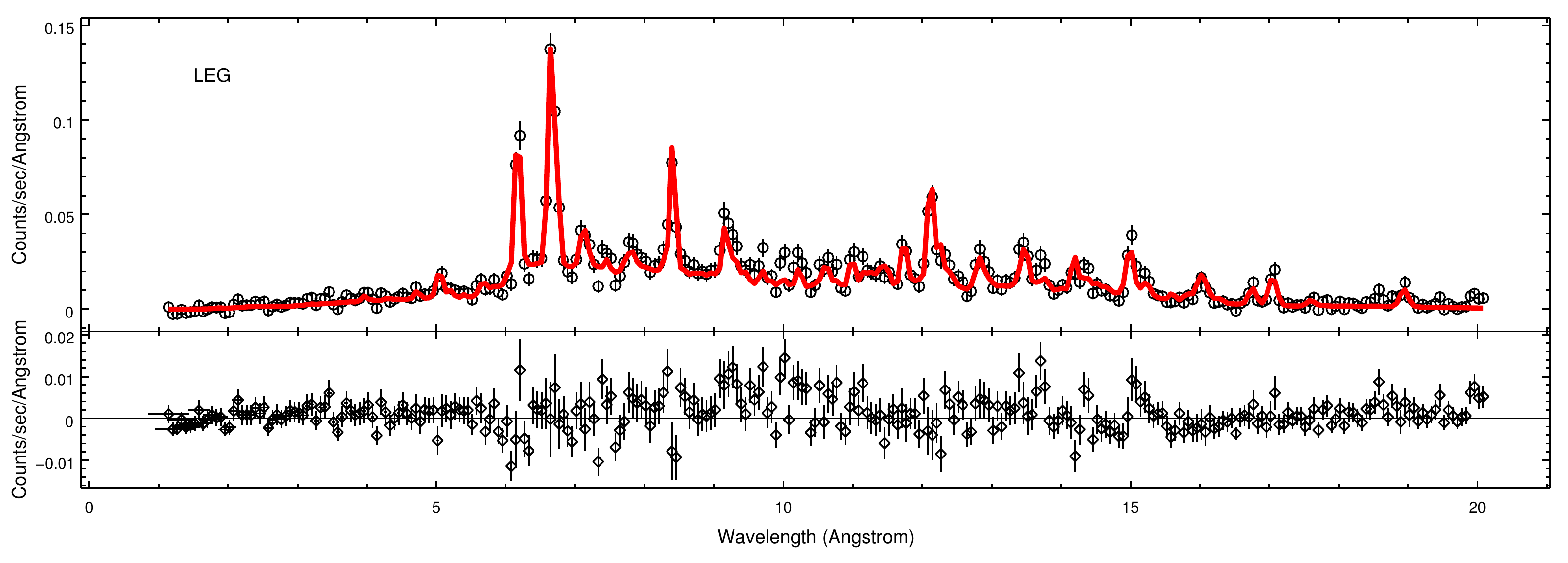}
\end{center}
\caption{The best-fit three-temperature model (see text) compared with the HEG (top) and MEG (middle) spectra binned by ten pixels, and similar best-fit LEG spectrum (bottom) model and data binned by 5 pixels. The lower sub-panels of each illustrate the fit residuals.
\label{f:fitting}}
\end{figure*}

\begin{figure*}
\begin{center}
\includegraphics[width=3.in]{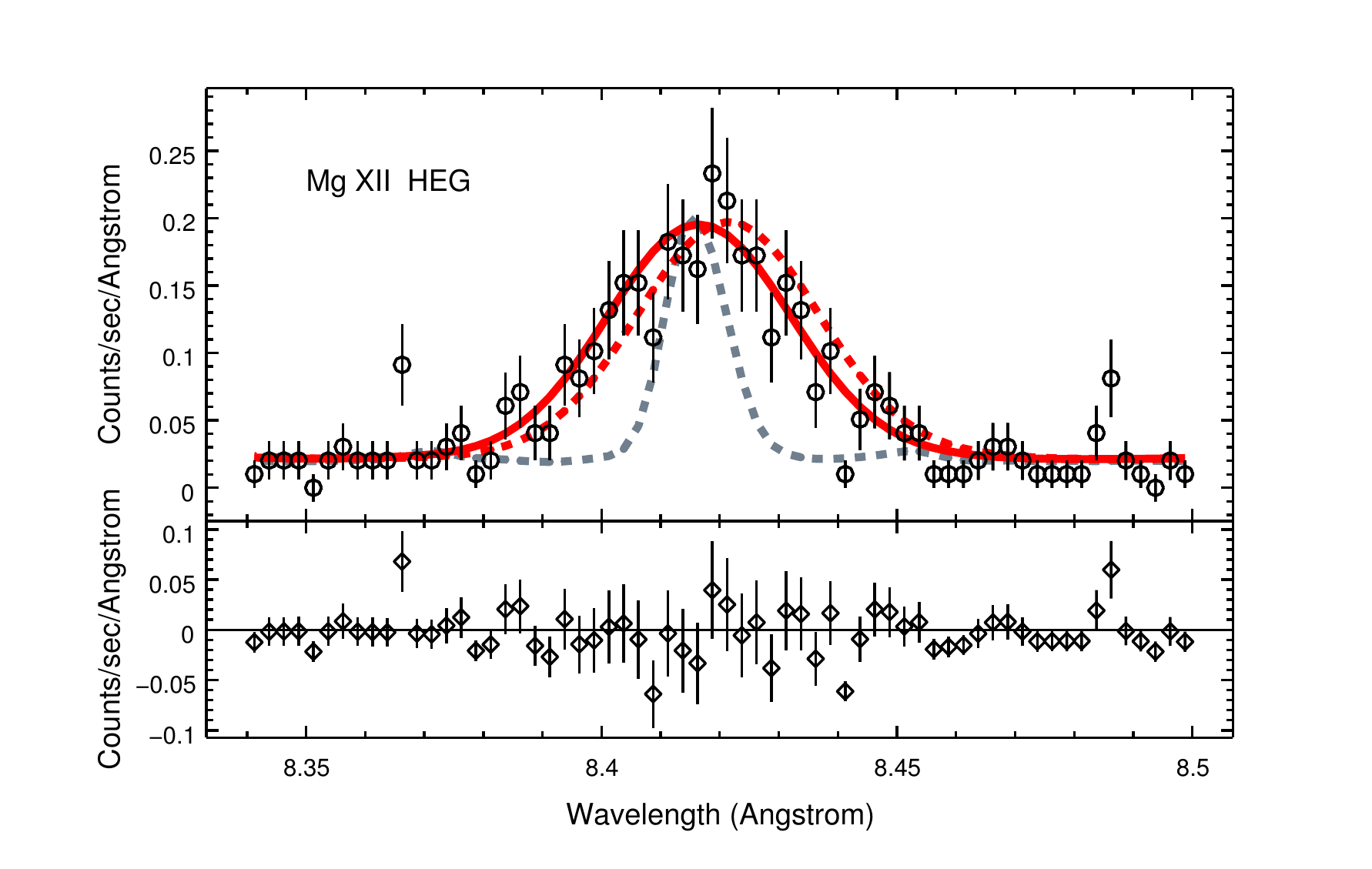} \includegraphics[width=3.in]{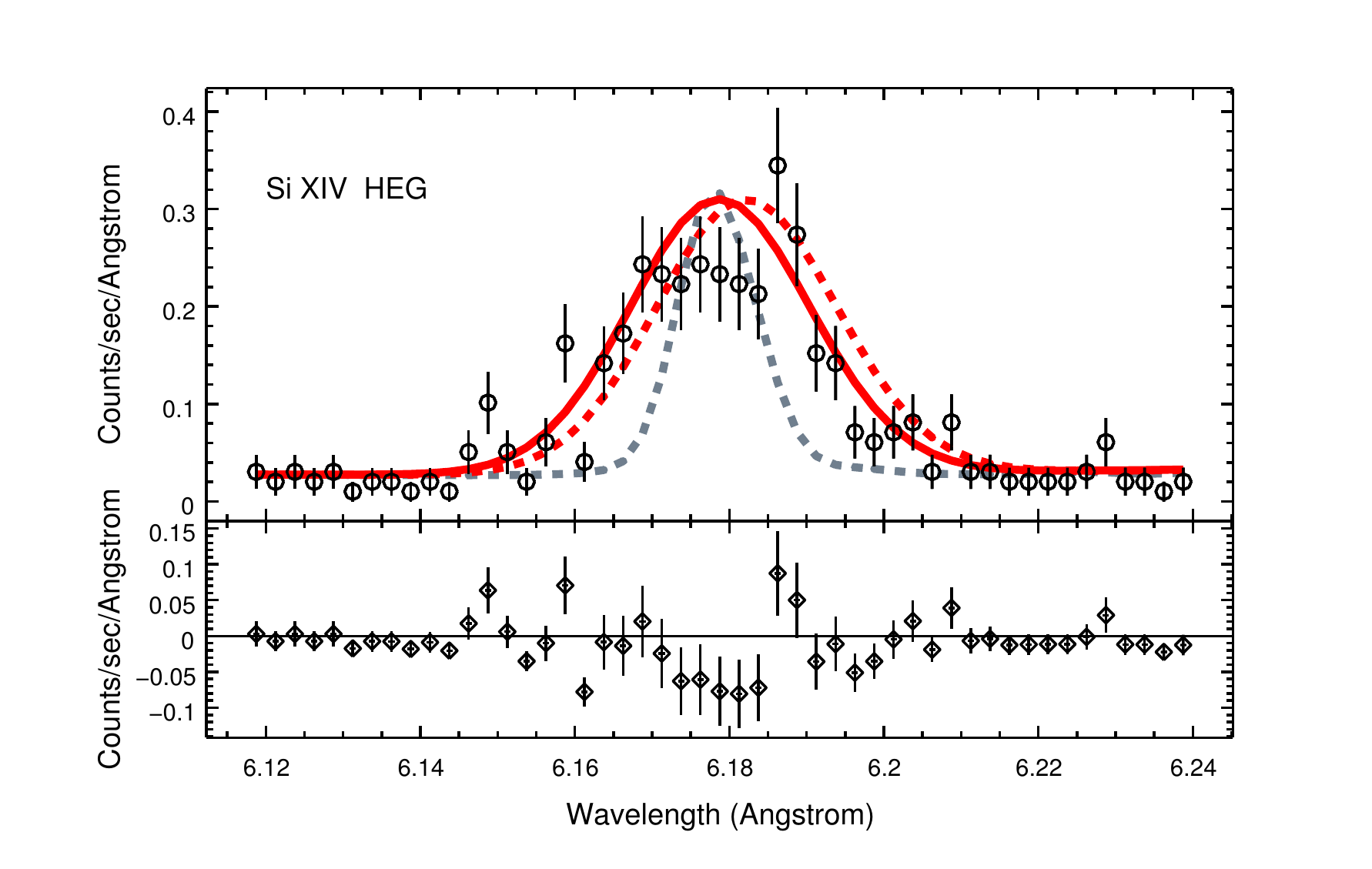}
\includegraphics[width=3.in]{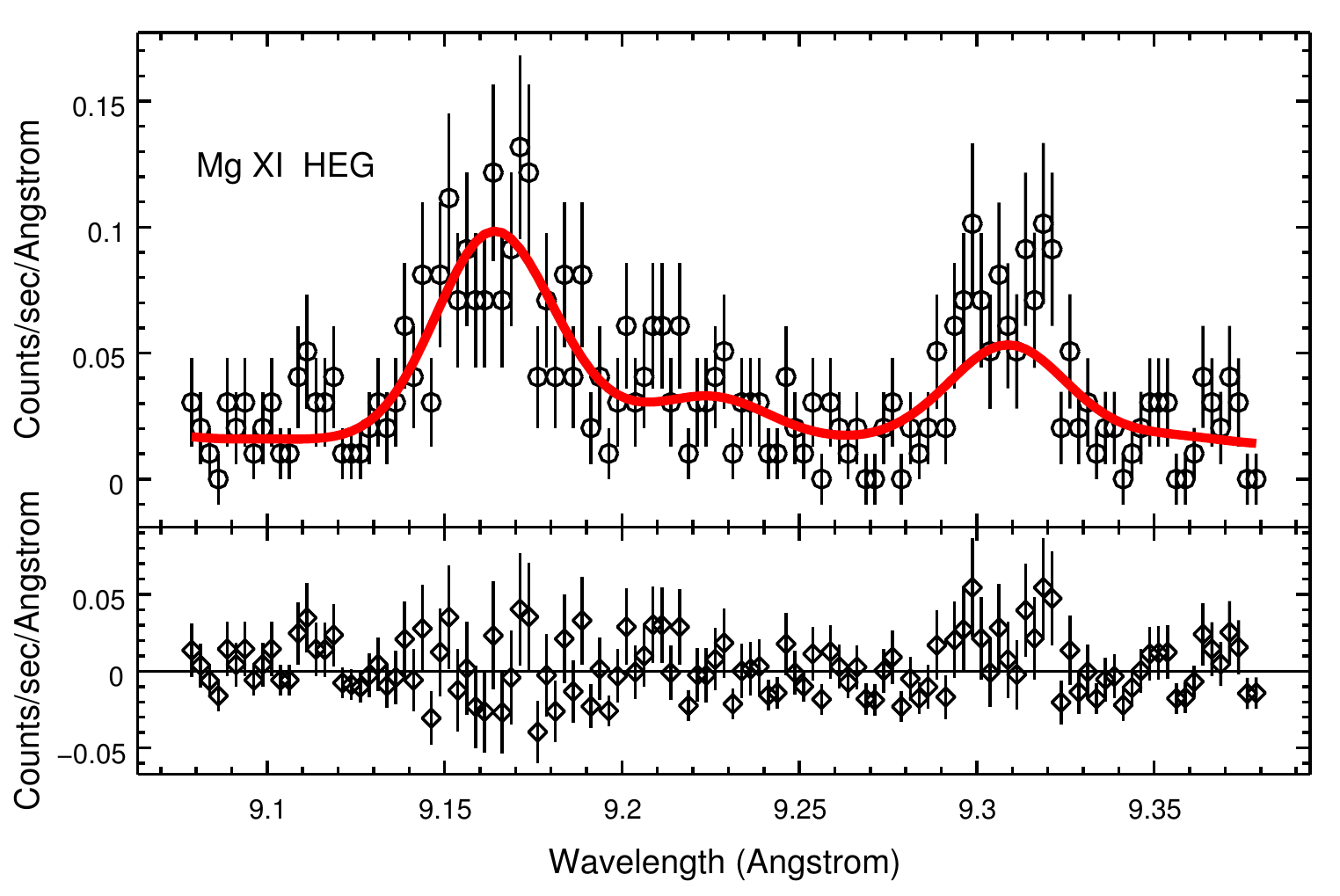} \includegraphics[width=3.in]{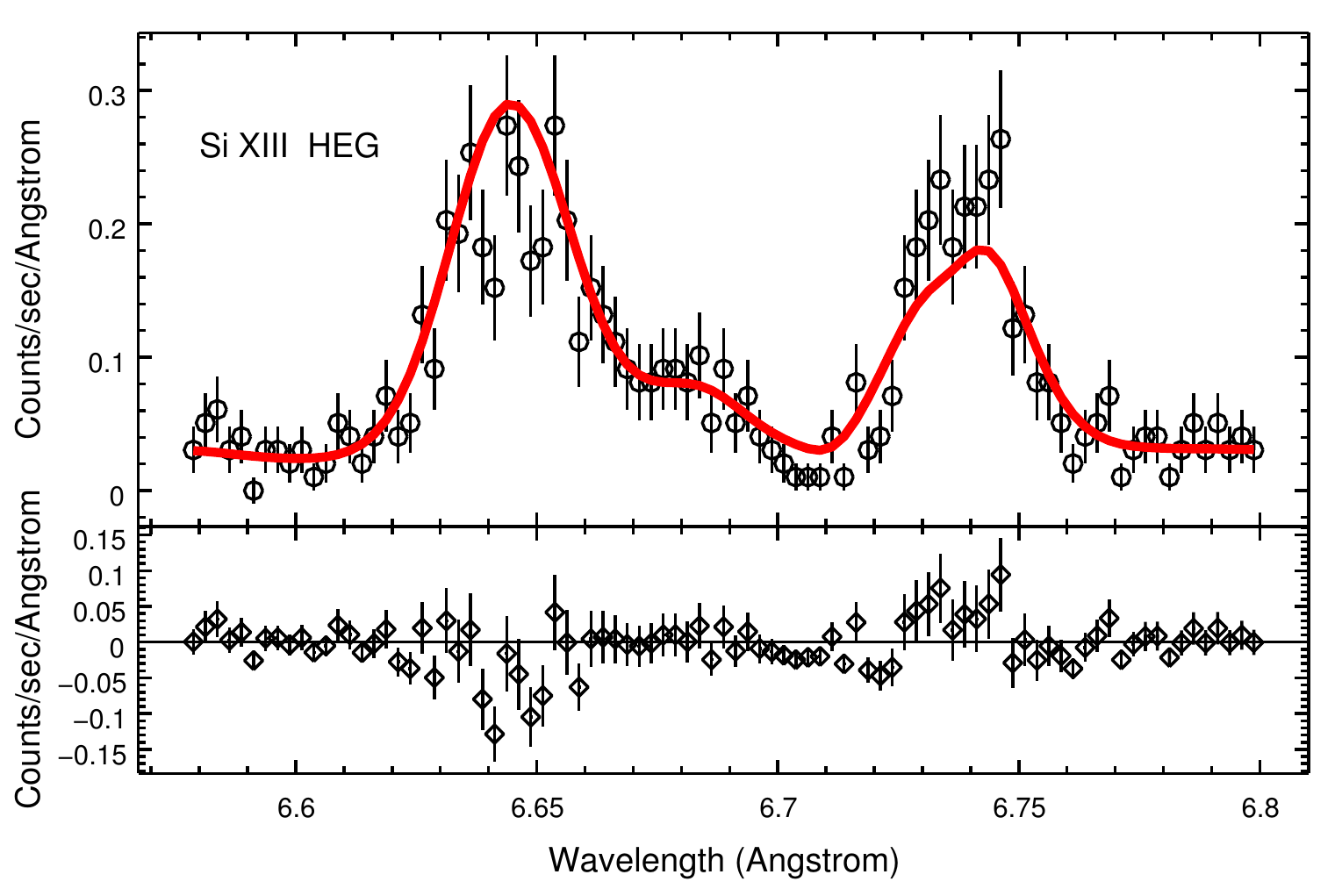}
\end{center}
\caption{The best-fit three-temperature model (see text) compared with the HEG spectra of the Mg (left) and Si (right) H-like (top) and He-like (bottom) lines.  The models include a velocity broadening of  1200~km~s$^{-1}$ FWHM and a blueshift of 165~km~s$^{-1}$. The dashed curves in the top panels correspond to the ``unshifted'' (redshift $z=0$; red dashed), and unbroadened (grey dashed) profiles.  The latter have been renormalized arbitrarily so as to have the same peak count rates as the others. 
In the lower panels, the three line components are the resonance (left; $1s^2\, ^1S_0$--$1s2p\, ^1P_1$), intercombination (center; $1s^2\, ^1S_0$--$1s 2p\, ^3P_2,^3P_1$), and forbidden (right; $1s^2\, ^1S_0$--$1s2s\, ^3S_1$) lines.  Their relative intensities correspond to the low-density limit (see Section~\ref{s:helike}).   \label{f:hhe}}
\end{figure*}

Adding further plasma components with normalizations and temperatures free to vary, but with abundances tied between components such that one set was common to all, further improved the HETG fit to $\mathcal C_r=1.09$ with two components and $\mathcal C_r=1.01$ with three.  Subsequent additional components resulted in only very marginal improvement in fit quality.  In the case of the LETG spectrum, with a reduced statistic close to unity, fits were already statistically acceptable and adding further thermal components only resulted in marginal gains in fit quality.  For a three-temperature, variable abundance model, the best-fit yielded $\mathcal C_r=0.70$.
The best-fit three-temperature models are compared with the full fitted range of the LEG, HEG and MEG observations, binned as a visual aid to highlight deviations between model and data, in Figure~\ref{f:fitting}, and with the H-like and He-like  profiles of Si and Mg unbinned and at full resolution in Figure~\ref{f:hhe}.   

While the abundance results are suggestive of a mild metal paucity with respect to solar values, and in this respect are in good agreement with the assessment of \citet{Orio.etal:15} based on {\it NuSTAR} and {\it Swift} observations, we caution against their strict interpretation.  The true abundance uncertainties can be significantly larger than the purely statistical errors owing to uncertainties and deficiencies in the plasma model, as well as in the instrument calibration (see also the discussion of the chemical composition of the ~RS~Oph blast wave emission by \citealt{Ness.etal:09}).  Moreover, as we note in Section~\ref{s:temp_dist} below, a multi-thermal solar metallicity model is not inconsistent with the data.  

\floattable
\begin{deluxetable}{l|rrcl}
 \tablecaption{HETGS spectral model parameter estimation results \label{t:hfit_results}}
 \tablehead{
\colhead{Model} & \colhead{Parameter\tablenotemark{1}} & \multicolumn{3}{c}{Best-fit value\tablenotemark{2}} \\
 }
\startdata
 & $N_H$ & (7.92 & $\pm$ & $0.10)\times 10^{21}$ \\
 Single-temperature,  & $kT$ & 1.135 & $\pm$ & 0.006\\
variable metallicity.   & $10^{[M/H]}$ & 0.46 & $\pm$ & 0.01 \\
Degs.\ of freedom:  9395 & Redshift & ($-59$ & $\pm$ & $_{3} ^4)\times 10^{-5}$ \\
Red.\ statistic:  1.33  & Velocity & 518 & $\pm$ & 11 \\
 & $EM$ & 0.0775 & $\pm$ & 0.0014 \\
  \hline
 & $\beta$ & $-0.67$ & $\pm$ & 0.13 \\
 & $kT_1$ & 0.603 & $\pm$ & $_{0.013} ^{0.012}$ \\
 & $EM_1$ & 0.0328 & $\pm$ & $_{0.0023} ^{0.0030}$ \\
 & $kT_2$ & 4.1 & $\pm$ & $_{0.6} ^{1.5}$ \\
 & $EM_2$ & 0.0070 & $\pm$ & $_{0.0017} ^{0.0015}$ \\
 Three-temperature,    & $kT_3$ & 1.21 & $\pm$ & $_{0.02} ^{0.03}$ \\
 variable abundances. & $EM_3$ & 0.0390 & $\pm$ & $_{0.0022} ^{0.0031}$ \\
 Degs.\ freedom:  9385  & O & 0.25 & $\pm$ & $_{0.08} ^{0.10}$ \\
Red.\ statistic:  1.01  & Ne & 0.64 & $\pm$ & $_{0.05} ^{0.06}$ \\
 & Mg & 0.62 & $\pm$ & $_{0.04} ^{0.03}$ \\
 & Al & 0.75 & $\pm$ & 0.09 \\
 & Si & 0.65 & $\pm$ & $_{0.04} ^{0.03}$ \\
 & S & 0.66 & $\pm$ & 0.04 \\
 & Fe & 0.26 & $\pm$ & $_{0.02} ^{0.03}$ \\
 & Redshift & ($-56$ & $\pm$ & $3)\times 10^{-5}$ \\
 & Velocity & 502 & $\pm$ & 10 \\
\hline
 & $\beta$ & 0.55 & $\pm$ & $_{0.04} ^{0.06}$ \\
& $EM_{6.4}$ & 0 & $\pm$ & $_{-} ^{0.0068}$ \\
& $EM_{6.6}$ & 0.0217 & $\pm$ & $_{0.0073} ^{0.0068}$ \\
 & $EM_{6.8}$ & 0.0204 & $\pm$ & $_{0.0045} ^{0.0032}$ \\
Eight-temperature,\tablenotemark{3} & $EM_{7.0}$ & 0.0282 & $\pm$ & $_{0.0026} ^{0.0014}$ \\
  variable metallicity& $EM_{7.2}$ & 0.0237 & $\pm$ & $_{0.0022} ^{0.0007}$ \\
 Degs.\ freedom:  9393  & $EM_{7.4}$ & 0 & $\pm$ & $_{-} ^{0.00087}$ \\
 Red.\ statistic:  1.02 & $EM_{7.6}$ & 0 & $\pm$ & $_{-} ^{0.0014}$ \\
 & $EM_{7.8}$ & 0.00287 & $\pm$ & $_{0.00088} ^{0.00017}$ \\
  & $10^{[M/H]}$ & 0.64 & $\pm$ & $_{0.01} ^{0.1}$ \\
 & Redshift & $(-56$ & $\pm$ & $_{3} ^{4})\times 10^{-5}$ \\
 & Velocity & 519 & $\pm$ & $_{11} ^{12}$ \\
\enddata
\tablenotetext{1}{Parameter units are: temperature, $kT$, in keV; emission measure , $EM$, in
$10^{-14}/ 4 \pi D^2$~cm$^{-3}$, where $D$ is the distance to the source; hydrogen column $N_H$ in cm$^{-2}$; redshift expressed in the usual ratio of velocity to that of the speed of light, $z=v/c$; broadening velocity in km~s$^{-1}$.
Abundances are expressed as a fraction of the solar abundances of \citet{Anders.Grevesse:89}; for element X this is equivalent to $10^{\rm [X/H]}$ in traditional spectroscopic notation.}
\tablenotetext{2}{Uncertainties correspond to $1\sigma$ (68.3\%) confidence bounds.}
\tablenotetext{3}{The eight temperatures are denoted by the subscripts to the emission measure in units of $\log_{10}(K)$.  In units of keV they are: 0.22, 0.34, 0.54, 0.86, 1.37, 2.16, 3.43, and 5.44~keV.}
\end{deluxetable}%

\floattable
\begin{deluxetable}{l|rrcl}
 \tablecaption{LETGS spectral model parameter estimation results \label{t:lfit_results}}
 \tablehead{
\colhead{Model} & \colhead{Parameter\tablenotemark{1}} & \multicolumn{3}{c}{Best-fit value\tablenotemark{2}} \\
 }
\startdata
 & $N_H$ & (8.74 & $\pm$ & $2.3)\times10^{21}$ \\
Single-temperature,  & $kT$ & 0.89 & $\pm$ & $_{0.18} ^{0.02}$ \\
variable metallicity.  & $EM$ & 0.1127 & $\pm$ & $_{0.0065} ^{0.0067}$ \\
Degs.\ of freedom:  725 & $10^{[M/H]}$ & 0.35 & $\pm$ & $_{0.03} ^{0.04}$ \\
 Red.\ statistic: 1.00 & Redshift & $-0.0019$ & $\pm$ & 0.0002\\
 & Velocity & 852 & $\pm$ & 78 \\
\hline
 & $\beta$ & $-0.57$ & $\pm$ & $_{0.40} ^{0.28}$ \\
 & $kT_1$ & 0.73 & $\pm$ & $_{0.71} ^{0.08}$ \\
 & $EM_1$ & 0.034 & $\pm$ & $_{0.033} ^{0.007}$ \\
 & $kT_2$ & 1.31 & $\pm$ & $_{1.29} ^{0.15}$ \\
 & $EM_2$ & 0.041 & $\pm$ & $_{0.039} ^{0.020}$ \\
Three temperature, & $kT_3$ & 5.44 & $\pm$ & $_{-} ^{-}$ \\
 variable abundances. & $EM_3$ & 0.0008 & $\pm$ & $_{-} ^{0.1}$ \\
  Degs.\ freedom:  718 & O & 0.27 & $\pm$ & $_{0.10} ^{0.16}$ \\
 Red.\ statistic: 1.01  & Ne & 0.69 & $\pm$ & $_{0.13} ^{0.24}$ \\
 & Mg & 0.61 & $\pm$ & $_{0.08} ^{0.22}$ \\
 & Al & 1.10 & $\pm$ & $_{0.39} ^{2.47}$ \\
 & Si & 0.64 & $\pm$ & $_{0.07} ^{0.60}$ \\
 & S & 0.47 & $\pm$ & $_{0.12} ^{0.46}$ \\
 & Fe & 0.26 & $\pm$ & $_{0.15} ^{0.11}$ \\
 & Redshift & $-0.0014$ & $\pm$ & 0.0002 \\
 & Velocity & 808 & $\pm$ & 88 \\
\hline
 & $\beta$ & 0.65 & $\pm$ & $_{0.10} ^{0.18}$ \\
 & $EM_{6.4}$ & 0.005 & $\pm$ & $_{-} ^{0.012}$ \\
 & $EM_{6.6}$ & 0.012 & $\pm$ & $_{0.012} ^{0.014}$ \\
 & $EM_{6.8}$ & 0.011 & $\pm$ & $_{0.008} ^{0.011}$ \\
Eight-temperature,\tablenotemark{3}  & $EM_{7.0}$ & 0.0308 & $\pm$ & $_{0.0068} ^{0.0075}$ \\
variable metallicity & $EM_{7.2}$ & 0.0213 & $\pm$ & $_{0.0028} ^{0.0082}$ \\
Degs. freedom: 722 & $EM_{7.4}$ & 0.0047 & $\pm$ & $_{-} ^{0.0006}$ \\
Red. statistic: 0.70 & $EM_{7.6}$ & 0 & $\pm$ & $_{-} ^{0.0029}$ \\
 & $EM_{7.8}$ & 0 & $\pm$ & $_{-} ^{0.0014}$ \\
 & $10^{[M/H]}$ & 0.74 & $\pm$ & $_{0.11} ^{0.22}$ \\
 & Redshift & $-0.00134$ & $\pm$ & $_{0.00024} ^{0.00022}$ \\
 & Velocity & 895 & $\pm$ & $_{79} ^{82}$ \\
\enddata
\tablenotetext{1}{Parameter units are the same as those in Table~\ref{t:hfit_results}.}
\tablenotetext{2}{Uncertainties correspond to $1\sigma$ (68.3\%) confidence bounds.}
\tablenotetext{3}{The eight temperatures are denoted by the subscripts to the emission measure in units of $\log_{10}(K)$.  In units of keV they are: 0.22, 0.34, 0.54, 0.86, 1.37, 2.16, 3.43, and 5.44~keV.}
\end{deluxetable}%

\subsubsection{Temperature distribution}
\label{s:temp_dist}

To probe the distribution of plasma temperature in the blast wave, an eight-temperature model was adopted, covering the temperature range 0.21--5.44~keV ($\log T =6.4$--7.8) in bins evenly distributed in logarithmic temperature.
As Tables~\ref{t:hfit_results} and \ref{t:lfit_results} testify, the three-temperature fits, in which abundances were allowed to vary, did not indicate drastic deviations from a solar abundance mixture, except for a mild paucity of metals, at 30--70\%\ of the solar values.  For the multi-thermal model, therefore, only the global metallicity was allowed to vary.  
While the formal $1\sigma$ uncertainty on the model metallicity appeared to preclude the solar value, test fits with the metallicity fixed at [M/H]$=0$ were found to result in only a very marginal increase in the test statistic, from $\mathcal C_r=1.02$ to 1.03.

\begin{figure*}
\begin{center}
\includegraphics[width=6.in]{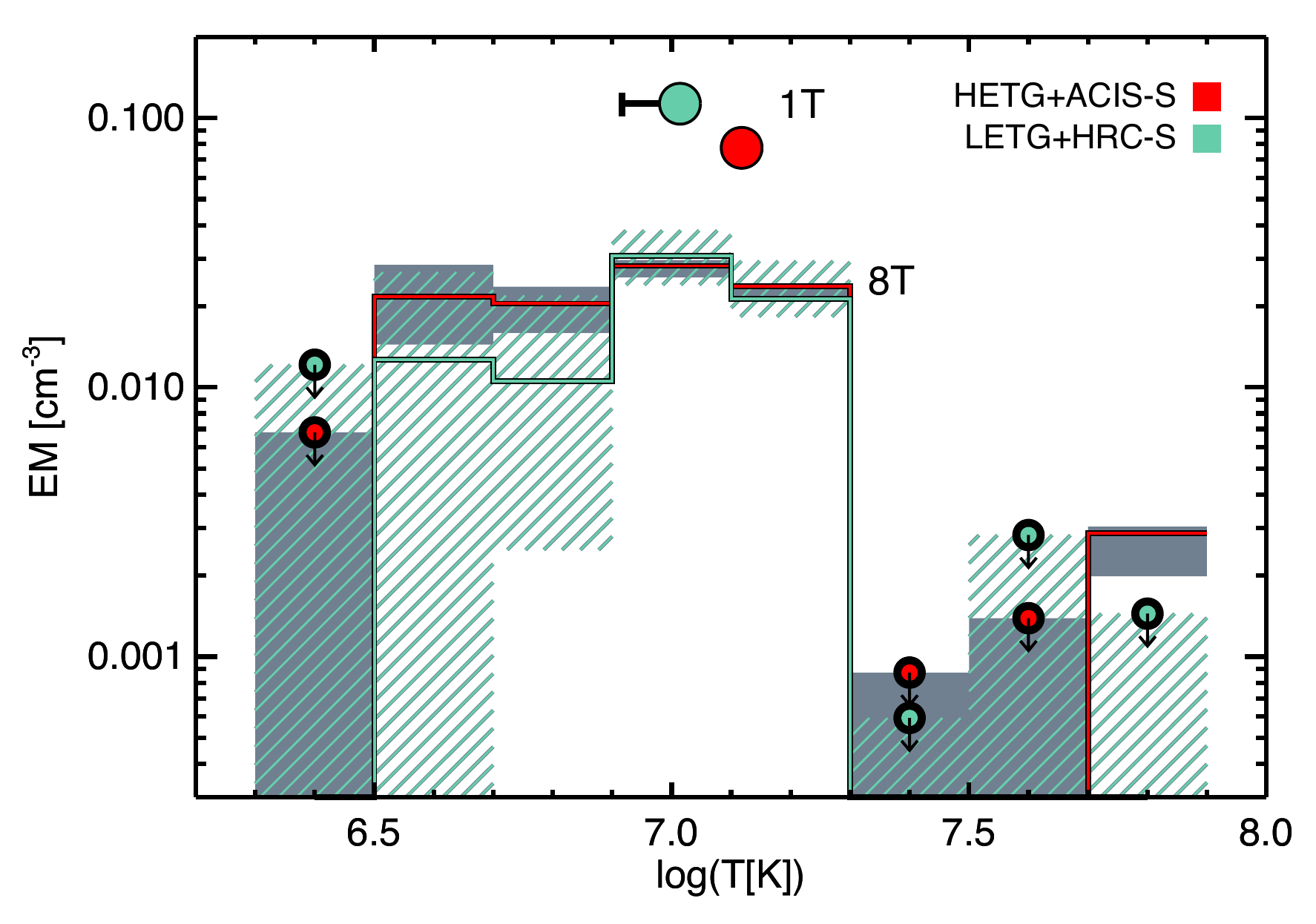}
\end{center}
\caption{The emission measure as a function of temperature for the eight-temperature model fit to HETG (HEG and MEG fitted simultaneously) and LETG spectra (histograms), together with the emission measures and temperatures from isothermal model fits (large round symbols).  The emission measures are in units of $10^{-14}/ 4 \pi D^2$~cm$^{-3}$, where $D$ is the distance to the source.  The shaded and hatched regions indicate the $1\sigma$ uncertainties at each temperature for HETG and LETG fits, respectively; $1\sigma$ upper limits on the multi-thermal histogram emission measures are indicated by downward-pointing arrows for temperature components that are not significantly present in the data.  Error bars on the isothermal model fits are smaller than the symbols, except for that shown.
\label{f:em}}
\end{figure*}

The best-fit emission measures as a function of temperature from the multi-thermal model are illustrated in Figure~\ref{f:em}, together with the single-temperature fit results for comparison. Several of the eight temperature bins were found not to have a significant emission measure and only upper limits could be derived.

Cooling flow models, in which the emission measure distribution follows the inverse of the radiative
loss function \citep{Mushotzky.Szymkowiak:88}, where also investigated.   Such models would be expected to give a reasonable description of
the X-ray spectrum for a one-dimensional steady-state, radiatively-cooled shock,
where the gas is heated to a single initial temperature.  In the case of V745~Sco, and as we discuss later in Sections~\ref{s:anal_blast} and \ref{s:colli}, the blast wave decelerates significantly, cools by adiabatic expansion as well as radiatively, and is likely to be more complicated than a one-dimensional model.   Unfortunately, cooling flow models as currently available in fitting engines also do not provide for ready inclusion of velocity broadening of spectral lines.  Nevertheless, we applied the {\it xsvmcflow} model within {\it Sherpa}, adopting the same abundance fitting strategy as employed for the three-temperature models described in Section~\ref{s:abuns}.  In addition to the abundances, the free parameters were maximum temperature, minimum temperature, normalization and redshift.  While a superficially reasonable fit was obtained, the reduced statistic $\mathcal C_r=1.6$ was formally unacceptable.  The best-fit minimum and maximum temperatures were 0.27 and 2.14~keV, respectively, and best-fit abundances were fairly uniformly close to $10^{[M/H]}\approx 0.4$, with the exception of Fe, for which  $10^{[M/H]}\approx 0.1$ was obtained.  While the poor statistical fit precludes reliable inference from the model, the temperature range found is very similar to the range of temperatures over which the emission measure was found to be significant from the eight-temperature model. 

\subsection{Velocity shift and broadening}

Fits of all models to HETG spectra, regardless of complexity, pointed to velocity broadening corresponding to a Gaussian $1 \sigma$ width of approximately 510~km~s$^{-1}$ and a redshift of -0.00056---actually a blueshift corresponding to $165 \pm 10$~km~s$^{-1}$.    Placing full weight on the three-temperature variable abundance and eight-temperature variable metallicity results, we find a velocity FWHM=$1200\pm 30$~km~s$^{-1}$, in good agreement with the \citet{Banerjee.etal:14} H~I line profile results. 

Unshifted and unbroadened profiles of H-like Mg and Si are also illustrated for reference in Figure~\ref{f:hhe}.  Owing to lower resolving  power, LETG spectra carry less velocity information.  At face value, both the best-fit broadening and blueshift obtained from the LETG data are significantly larger than found from the HETG.  However, the HRC-S detector is known to suffer from low-level dispersion non-linearities\footnote{http://cxc.harvard.edu/proposer/POG/}
 that render fine Doppler shift analysis questionable; we therefore do not consider the LETG velocity results further.

\subsubsection{He-like Ions}
\label{s:helike}

The so-called ``triplets" of helium-like ions are well-known as density and temperature diagnostics for plasmas in collisional equilibrium (\citealt{Gabriel.Jordan:69}; see also \citealt{Porquet.etal:10} for a recent review).  Based on the shock velocity evolution estimated by \citet[][see also Section~\ref{s:anal_blast} below]{Banerjee.etal:14}, the shock radius is expected to be roughly 15~AU, and the shocked circumstellar medium should have a density not much higher than $n_e\sim 10^8$~cm$^{-3}$.  This is orders of magnitude below the lower limits of density sensitivity of the He-like Ne, Mg and Si lines, which are a few $10^{10}$, $10^{11}$, and $10^{12}$~cm$^{-3}$, respectively \citep[see, e.g.,][]{Porquet.etal:10}.   

The observed forbidden and intercombination lines illustrated in Figure~\ref{f:hhe} are reasonably consistent with the predicted low density limit intensities, consistent with the estimated shocked circumstellar medium density expectations noted above.  At higher densities above the low density limit of sensitivity, the metastable $^3S_1$ upper level of the forbidden line is collisionally excited to $^3P_{0,1,2}$ at a sufficient rate to increase the intercombination line strength at the expense of the forbidden line.   Similarly, the $^3P_{0,1,2}$ levels can be excited radiatively in the presence of a strong UV radiation field, which is lacking in the post-SSS phase of V745~Sco.
We do note, though, a tendency for the forbidden line to appear slightly under-predicted by the model.  The shocked circumstellar medium in the V745~Sco blast wave is in a cooling state and therefore expected to be a recombining plasma.  Since this process favors the larger statistical weight of the triplet over the singlet levels during recombination cascades, the forbidden line can be enhanced relative to its strength in the pure collisional equilibrium state.

\subsection{Inference from spectral line profiles}
\label{s:profiles}

The spectral line profiles of V745~Sco bear superficial resemblance to the profiles observed for the blast wave of the recurrent nova RS~Oph (see \citealt{Drake.etal:09}), with an increasing apparent blueshift of line centroids going from shorter to longer wavelengths.  This pattern is a signature of intrinsic absorption within the remnant that eats away the far side emission producing the red-shifted portion of the line.  The effect is most striking for the H-like O~VIII doublet at 18.97~\AA.  While the signal-to-noise ratio of this line is very low, with only a handful of photon counts, nearly all of these counts lie to the blue of the expected line centroid.

\begin{figure*}
\begin{center}
\includegraphics[angle=0,width=6.2in]{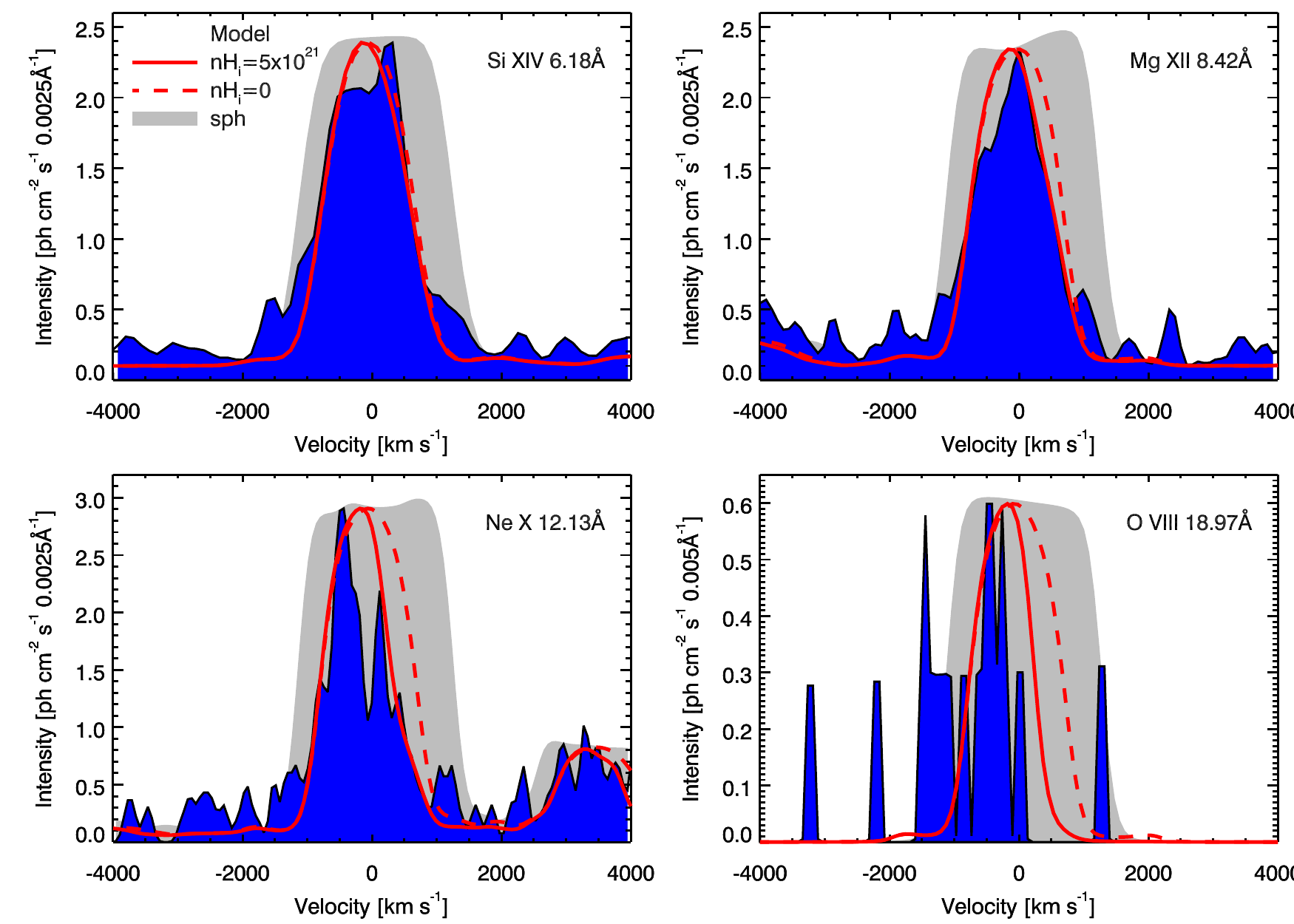}
\end{center}
\caption{Observed velocity profiles of the H-like resonance lines of Si {\sc xiv}, Mg {\sc xii}, Ne {\sc x}, and O {\sc viii} (shaded blue) compared with simple emission models.  Solid grey profiles represent an unabsorbed thin spherical shell with an expansion velocity of 1200~km~s$^{-1}$; dashed red profiles correspond to a spherical cap with an opening angle of $50^\circ$ inclined toward the observer at $5^\circ$ relative to the plane of the sky and expanding with a velocity of 2000~km~s$^{-1}$; solid red curves correspond to the same model subject to intrinsic absorption interior to the sphere with an equivalent hydrogen column density of $N_H=5\times 10^{21}$~cm$^{-2}$.
\label{f:velocity}}
\end{figure*}

The pointed shape of spectral lines, combined with the net blueshift, even for profiles at short wavelengths relatively unaffected by the differential intrinsic absorption, indicates a highly collimated, rather than spherically-symmetric, blast.   Figure~\ref{f:velocity} illustrates the profiles of the Ly$\alpha$ transitions in H-like Si, Mg, Ne and O compared with idealized model emission profiles.  The latter correspond to emission from an expanding thin spherical shell, for which we can vary the expansion velocity, the ranges of latitude that contribute to the emission, the inclination of the reference axis of the shell to the plane of the sky, and the absorbing column within the shell.  These model emission profiles were convolved with the best-fit eight-temperature emission model described in Section~\ref{f:fitting} and with the instrumental profiles of the HEG and MEG grating spectrometers before scaling to match the observed peak profiles for comparison.  

Square, pedestal-like spectral line profiles are sometimes observed in connection with nova explosions; such profiles result from a quasi-spherical expansion of emitting material.  Line profiles of V745~Sco correspond to a full width at zero intensity (FWZI) of approximately 2400~km~s$^{-1}$, corresponding to a maximum expansion velocity of the X-ray emitting gas of  $v_{exp}=1200$~km~s$^{-1}$.  The observed line shapes are much more peaked than those from spherical shell-like emission with such an expansion velocity.  The net blueshift, even in Si~XIV that suffers minimal intrinsic absorption, is also conspicuous.  

Different configurations of spherical shell sections with $v_{exp}=1200$~km~s$^{-1}$ were also qualitatively compared with the observed line profiles.  Lacking a detailed spatial and physical emission model, no attempt was made to obtain best-fit model parameters: the profiles merely serve to demonstrate that a collimated blast model can provide a reasonable match to the data, even though they fail to match them in detail.
Both ring-like (emission restricted to an equatorial belt below a specified latitude, $\phi$) and cap-like (emission restricted to one pole, above a specified $\phi$) models were found to produce profiles similar to those observed.  Both require dominant emission from material expanding predominantly in the plane of the sky.  In the case of ring-like emission, a match was found for an emitting region of $-15^\circ \leq \phi \leq 15^\circ$ with a system 
inclination $i\sim 25^\circ$ (i.e.\ the axis of the ring inclined at $65^\circ$ relative to the sky plane).  Alternatively, a spherical cap with $\phi \geq 65^\circ$ and $i\sim 85^\circ$ was also found to provide a good qualitative profile match.  Two sets of line profiles corresponding to this latter model are illustrated in Figure~\ref{f:velocity}; one includes the effect of intrinsic absorption within the spherical cavity while the other is unabsorbed.  An equivalent neutral hydrogen column density of $5\times 10^{21}$~cm$^{-2}$ within the cavity was found to provide a reasonable match to the differential absorption across the different line profiles, with, perhaps, the exception of the poorly-detected O~VIII line.

\subsection{{\it Swift} XRT Observations}
\label{s:swift_anal}

{\it Swift} XRT observations were analyzed following the procedure outlined by \citet{Page.etal:15}.  We are mainly interested in using the {\it Swift} data to follow the temperature of the blast wave as a function of time.  Model parameter estimation was undertaken using the {\sc XSPEC} fitting engine, applying isothermal APEC models attenuated by a variable absorption component representing the circumstellar environment combined with a fixed interstellar absorption corresponding to a neutral hydrogen column density of $5.6\times 10^{21}$~cm$^{-2}$.  A metallicity of 0.51 times the solar abundances of \citet{Anders.Grevesse:89} was adopted, following \citet{Orio.etal:15}.  Unfortunately, during the SSS phase there was insufficient signal at higher energies above that of the SSS signal with which to constrain the plasma temperature; consequently there is a gap in the {\it Swift} temperatures between days 4-12. We discuss the best-fit temperatures in Section~\ref{s:anal_blast}, below.

\section{ANALYTICAL BLAST WAVE MODEL}
\label{s:anal_blast}

In order to glean further insights into the explosion parameters, we follow the 
blast evolution using an
analytic model developed for supernova remnants (SNRs) by \citet{Laming.Hwang:03} and \citet{Hwang.Laming:12}.  This approach implements self-similar
models with an assumed ejecta density distribution comprising a uniform
density core surrounded by a power-law outer envelope. The forward
and reverse shock velocities are calculated as a function of
time.  

Assuming no differential absorption across the nova remnant, the FWHM of
emission lines should be $1.8v_s$, where $v_s$ is the shock velocity.  The thermal width of shocked
ions is approximately $v_s$, which is added in quadrature with $1.5v_s$, the difference in bulk
velocity between the front and back portions of the shock. 
We take a circumstellar medium consistent with that adopted by \citet{Banerjee.etal:14}, with density profile $\rho =
0.1r_{pc}^{-2}$~a.m.u.~cm$^{-3}$, where $r_{pc}$ is the radial distance in parsecs. This corresponds to a pre-nova mass loss rate of about
$\dot{M}_{wind}=3\times 10^{-7} M_{\sun}$ yr$^{-1}$ with a terminal wind speed of $v_{wind}=10$~km
s$^{-1}$.  Again following \citet{Banerjee.etal:14}, we take an ejected mass of $10^{-7} M_{\sun}$, and assume an 
outer envelope power law index of $-12$ for the ejecta \citep[see][]{Laming.Hwang:03}.  The ejected mass combined with the early Pa$\beta$ line widths of about 5000~km~s$^{-1}$ implies an initial expansion velocity of approximately 3000~km~s$^{-1}$ and an explosion energy of the order of $10^{43}$~erg; we take these parameters as our reference case.  Further, we explore a model with an ejected mass of $3\times 10^{-7}M_\odot$, one with an explosion energy $3\times 10^{43}$~ergs, 
 and one with a $1/r$ density profile, in addition to exploratory calculations with a uniform circumstellar density and yet higher ejected masses.

\subsection{Comparison with Near-Infrared Line Widths}

\begin{figure*}
\begin{center}
\includegraphics[angle=0,width=6.2in]{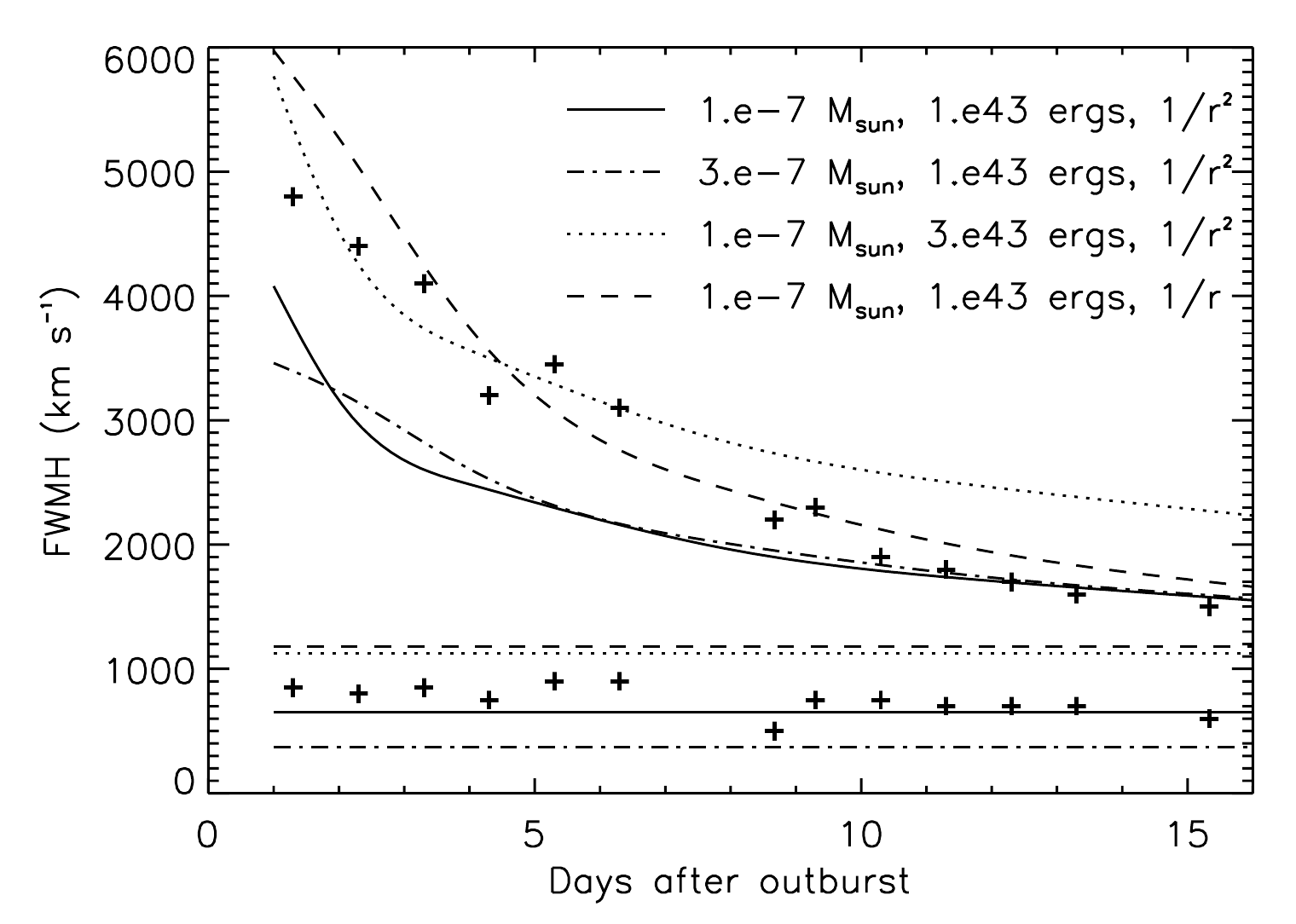}
\end{center}
\caption{Predicted H line widths for our analytical blast wave models (curves) compared with observed H~I~Pa$\beta$ widths from \citet[][crosses]{Banerjee.etal:14}.  The upper curves and data points correspond to the broader line component arising in the forward shock; the lower set of curves in our models arise from the reverse shock which we identify with the narrow component in the line widths of \citet{Banerjee.etal:14}.  Four models are shown, corresponding to different combinations of ejected mass, explosion energy and circumstellar density profile (see text). 
\label{f:linewidths}}
\end{figure*}

Figure~\ref{f:linewidths} reproduces the data presented 
in Figure~3 of \citet{Banerjee.etal:14}, who observed the profile
of the H~I Paschen $\beta$ emission line ($n=5$ to $n=3$) between one and
sixteen days after outburst. The profile is seen to be composed of
broad and narrow components. The broad component, which we interpret
as coming from the forward shock, declines from about 5000~km~s$^{-1}$
FWHM at early times to about 1500~km~s$^{-1}$ by day 16. The narrow
component, interpreted here as coming from the reverse shock, is
approximately constant with a value of about 700~km~s$^{-1}$.  We also show
four sets of curves corresponding to the different models described above.  Each set comprises the curves corresponding to reverse and forward shocks.  Unfortunately, no measurement uncertainties are quoted in \citet{Banerjee.etal:14}.
 The scatter  in the measurements about smooth trends indicates that these are of the order of 10\%.

None of the models are able to reproduce the exact trends of both broad and narrow line components throughout the time interval studied.  In all 
models, the reverse shock is propagating through the ejecta core 
with an approximately constant velocity, as is reflected by the constant line width of the shocked ejecta shown in Figure~\ref{f:linewidths}.
Curves corresponding to the reference case, with an explosion energy of
$10^{43}$ ergs and ejected mass $10^{-7}M_\odot$,  give the better match to the narrow component, and also match the broad component line widths after day 8.
Earlier in the evolution, near day 1,
the radius of the shock wave is similar to the putative separation of the
binary components in V745 Sco, and the difference in the center of the
$1/r^2$ density profile due to the wind from the companion, and the
center of the explosion on the white dwarf (assumed zero in the model),
should be expected to lead to inaccuracies.

The model with a $1/r$ circumstellar density profile also provides a reasonable match to the observed broad component line widths, with small overestimates at days 1 and 2.   A constant 
density profile greatly overestimates the line widths at early times and was not considered further.  The problem with these profiles is that they lead to 
reverse shock speeds that are too high and the predicted narrow component widths are too broad. With $n \sim 1/r$, this can be fixed by adopting a very steep outer ejecta envelope,  $n=20$ for example, although such a dependence is difficult to justify.  For a constant density medium, it is just not possible to obtain a reverse shock that matches the observations. 

Larger ejecta masses in the $n\sim 1/r^2$ model were also examined, but these produced smaller line widths at early times, and worse agreement with
observations.

\subsection{Comparison with X-ray Observations}
\label{s:xcomp}

\begin{figure*}
\begin{center}
\includegraphics[angle=0,width=6.2in]{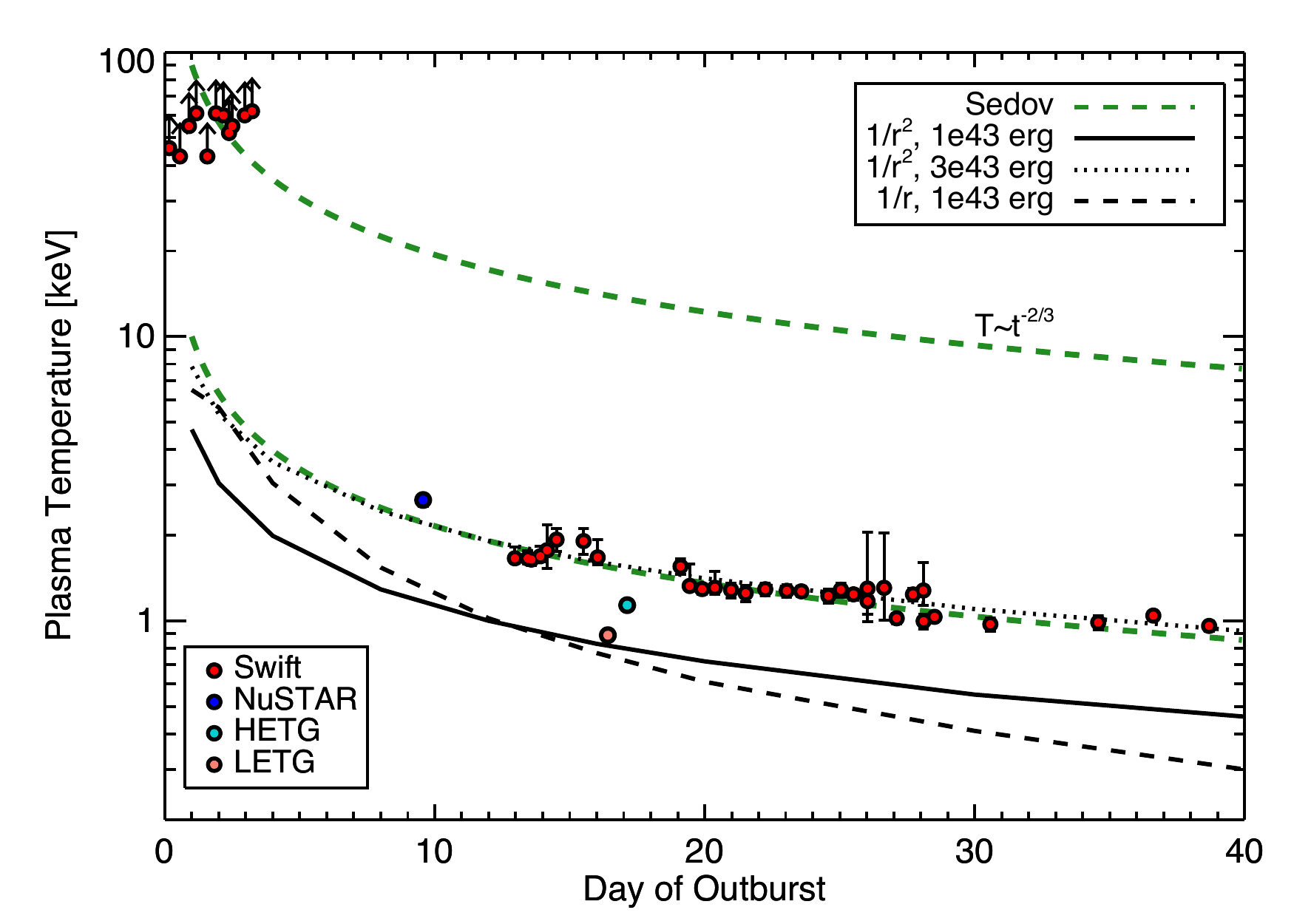}
\end{center}
\caption{The plasma temperature vs.\ time for the V745~Sco blast wave estimated from {\it Swift} XRT data (see Section~\ref{s:swift_anal}), together with that estimated from the {\it NuSTAR} observation by \citet{Orio.etal:15}.   The electron  temperature evolution of three of our analytical blast wave models, together with the expected temperature evolution based the Sedov relations (see text), are also shown.
\label{f:swift_kt}}
\end{figure*}

We compare the electron temperatures of the analytical models as a function of time with X-ray observations in Figure~\ref{f:swift_kt}.  There are two striking features of these comparisons.  Firstly, the early-time {\it Swift} temperatures are an order of magnitude higher than predicted by {\em any} of our models---no reasonable combination of model parameters can be chosen to reach such high initial temperatures.  These temperatures are also inconsistent with the early-time broad component line widths of \citet{Banerjee.etal:14}. Secondly, the later-time trend of temperature with time is extremely well-matched both by a Sedov decay law (see below) and by our models when a $1/r^2$ circumstellar density profile is adopted.

A  maximum electron temperature of about $9\times 10^6$~K is 
reached in the reference $10^{43}$~erg model on days 16--17.  This is lower by about a factor of 2 than the observed 
trend in the {\it Swift} temperatures.  The {\it Chandra} temperatures are significantly lower than those derived using {\it Swift} data, highlighting the instrument-dependent systematic uncertainty involved in assigning an average temperature to what is more of a continuous distribution of temperatures.
The reference model modified with an explosion energy of $3\times 10^{43}$~ergs slightly over-predicts the observed line widths in Figure~\ref{f:linewidths} but provides an excellent match to the {\it Swift} temperatures.

A Sedov-type blast (instantaneous release of energy at a point; \citealt{Sedov:59}) into a $\rho \propto r^{-w}$ density gradient leads to the shock wave radius, $r_s(t)$, expansion law with time $r_s \propto t^{2/(5-w)}$.  For an inverse square density law, $r_s\propto t^{2/3}$ and the shock velocity decreases as $v_s\propto t^{-1/3}$.  The temperature therefore evolves with time as $T\propto t^{-2/3}$.  This relation is also illustrated in Figure~\ref{f:swift_kt} and provides an excellent match to the observed temperature evolution.

The difficulties in matching simultaneously both observed spectral line widths and plasma temperatures using what are fairly simple one-dimensional models points to complicating factors in the blast that are not included in the models.  We discuss these below in Section~\ref{s:discuss}.  Nevertheless, we consider the general agreement between our reference and $3\times 10^{43}$~erg energy models  and the optical and X-ray data as confirmation that both the explosion energy of 1--$3\times 10^{43}$~erg and ejecta mass of $10^{-7}M_\odot$ are roughly correct.

The assessment of the shocked gas emission measure and temperature afforded by the high-quality {\it Chandra} spectra  provides an additional test of our blast wave model.  The best-fit single temperature plasma model from the HETG spectral analysis, which we deem more accurate than the LETG-derived  parameters for the reasons noted in Section~\ref{s:fitting}, has a temperature $kT=1.13$~keV and emission measure $7.8\times 10^{-16}/4\pi D^2$~cm$^{-3}$.   The uncertain distance to V745~Sco poses a potential problem for comparisons.  \citet{Mroz.etal:14} discount the distance of 7.8~kpc derived by \citet{Schaefer:10} because their extensive OGLE photometry rule out the period \citet{Schaefer:10} used to constrain the red giant radius.  While this is a reasonable conclusion, \citet{Schaefer:10} note that several earlier studies place the star in the Galactic Bulge giant population. Since it is unlikely that it lies behind the bulge, and the consensus distance to the Galactic center has long since converged at approximately 8~kpc \citep[e.g.][]{Reid:93}, 7.8~kpc remains a reasonable distance and for consistency with earlier studies we adopt it here.  We then find an emission measure of  $EM=5.7\times 10^{58}$~cm$^{-3}$.

The blast wave model density is $\rho = 0.1$~a.m.u.~cm$^{-3}/r_{pc}^2$, so the total swept up mass is 
\begin{equation}
M_{sw}=\int_0^{r_{pc}} 4\pi\rho(r) r^2\, dr =  0.1\times (3.1\times 10^{18})^2\, r_{pc} \,\, {\rm a.m.u.}, 
\label{e:msw}
\end{equation}
which gives $1.20\times 10^{37} r_{pc}$.  For the two models illustrated in Figure~\ref{f:linewidths}, the forward shock radius at day 17 and the time of the {\it Chandra} HETG observations is $r_{pc}=6.0\times 10^{-5}$ and $8.7\times 10^{-5}$ (12.4 and 17.9~AU)
for the two models, with the larger radius corresponding to the more energetic explosion.  The swept up mass is then  $M_{sw}=2.28\times 10^{51}$ and $3.24 \times 10^{51}$~a.m.u.   The emission measure is $EM=4\rho M_{sw}$ for strong shock conditions (assuming post-shock compression by a factor of four and that the shock structure is not significantly modified by particle acceleration; \citealt{Tatischeff.Hernanz:07}; see also Section~\ref{s:accel}). The density at $r_{pc}$ is $\rho=2.6\times 10^7$ and $1.3 \times 10^7$~cm$^{-3}$ so the emission measure is $EM =1.7 \times 10^{59}$ and $2.4\times 10^{59}$~cm$^{-3}$,  with the lower value coming from the higher energy model.   Given the remaining uncertainties in the distance, and the fact that we are comparing a spherically-symmetric model to what we have demonstrated to be a non-spherical blast wave, we consider the predicted and observed emission measures to be in good agreement.  The theoretical overestimate of the emission measure is plausibly due to our neglect of shocked plasma cooling out of the relevant temperature range.

{\it NuSTAR} observed the V745~Sco blast 10 days after discovery \citep{Orio.etal:15}---approximately a week before the {\it Chandra} campaign reported here.  \citet{Orio.etal:15} estimated a plasma temperature $kT = 2.66\pm ^{0.09}_{0.14}$~keV based on the {\it NuSTAR} data alone, which is significantly higher than the value $kT = 1.135\pm 0.006$~keV we obtain from  isothermal fits to the HETG spectra, and is also higher than later {\it Swift} temperatures.  While we find multi-thermal fits do reveal evidence for the presence of significantly hotter plasma---our three-temperature variable abundance fits resulted in only a weak hot temperature component of $kT = 4.1\pm ^{1.5}_{0.06}$~keV---the indication is that the bulk of the X-ray emitting plasma was significantly cooler on day 17 than on day 10.  

The shocked gas cools by radiation and adiabatic expansion, with our blast wave model indicating the shocked gas cooled in about a day at post-outburst times of a day or so.  The radiative cooling time is 
\begin{equation}
\tau_R=\frac{3kT}{n_e\Lambda(T)},
\end{equation}
where $\Lambda(T)$ is the radiative loss per unit emission measure.  We have calculated the expected radiative loss using the CHIANTI atomic database version 7.1.3 \citep{Dere.etal:97,Landi.etal:13} as implemented in the PINTofALE\footnote{PINTofALE is freely available at http://hea-www.harvard.edu/PINTofALE} IDL-based software suite \citep{Kashyap.Drake:00} and find $\Lambda(T)\sim 3 \times 10^{-23}$ erg~cm$^3$~s$^{-1}$ for a plasma at $10^7$~K with a metallicity $10^{[M/H]}$=0.46, so that the radiative cooling time is $\tau_R=1.4\times 10^{14}/n_e$.  The adiabatic expansion cooling time is approximately $\tau_A=r_s/v_s$, and both radiative and adiabatic cooling times increase with time as the shock radius increases and the velocity and density decrease.  
The adiabatic cooling time is then approximately $2\times 10^6$~s for a shock velocity of $v_s\sim 1000$~km~s$^{-1}$ around day 17.  The circumstellar density is $\rho=\dot{M}/4 \pi r_s^2 v_{wind}$, which, for the average of our models, amounts to $n_e=2 \times 10^{7}$~cm$^{-3}$.  Applying strong shock jump conditions as above in Equation~\ref{e:msw}, the post-shock gas density will be four times this, so that the radiative cooling time is $\tau_R\sim 2\times 10^6$~s---the same as the adiabatic cooling time for a combined cooling time of about 12 days.   

Since radiative cooling for plasma at $10^7$~K depends primarily on the density, the circumstellar density law and the Sedov radius dependence on time can be combined to find that the radiative cooling time scales with time as $\tau_R\propto t^{4/3}$; similarly the adiabatic cooling time scales as $\tau_A\sim t$, such that at very late times adiabatic cooling will dominate.   The {\it NuSTAR} observations occurred close to a week before the {\it Chandra} observations when the cooling time would have been about 6 days.  The shocked medium observed by {\it NuSTAR} would then have cooled significantly by the time of the {\it Chandra} observations, consistent with our finding of lower temperatures from those data.

The multi-thermal eight-temperature model revealed the presence of plasma at temperatures all the way from $3\times 10^6$ to $40\times 10^6$~K.  The lower temperatures result from the cooling of plasma that was shock-heated several days earlier in the history of the blast.

\section{DISCUSSION}
\label{s:discuss}

\subsection{Collimation and Asymmetry}
\label{s:colli}

{\it Chandra} high resolution spectroscopy of the V745~Sco explosion reveals clear evidence of a collimated blast.  Line profiles are both much more pointed than the square-shaped, boxy profiles expected from a spherically-symmetric blast, and significantly blue-shifted.  The FWZI of approximately 1200~km~s$^{-1}$ is to be compared with the smaller net blueshift of 165~km~s$^{-1}$; this latter figure is likely slightly biased toward larger blue shift by the differential absorption discussed in Section~\ref{s:profiles}, although we point to Figure~\ref{f:hhe} as an illustration that this does also seem appropriate for the relatively unabsorbed Si and Mg H-like and He-like resonance lines. 
Combined, these characteristics point to emission from material expanding preferentially in a direction close to the plane of the sky. 

As noted in Section~\ref{s:intro}, there is a growing body of evidence pointing to collimation and asymmetry of nova explosions, a topic that might be considered as beginning with the detailed discussion of spectral line profiles of novae by \citet{Hutchings:72}.  In addition to the radio and HST images of RS~Oph \citep{Hjellming.etal:86,Taylor.etal:89,O'Brien.etal:06,Bode.etal:07,Rupen.etal:08,Sokoloski.etal:08}---the most studied nova blast wave to date---observations of other novae have shown similar characteristics.  Examples include infrared interferometry of V1663~Aql revealing asymmetry \citep{Lane.etal:07}, both radio and optical 
 HST observations of light echoes revealing an asymmetric remnant of T~Pyx outbursts \citep{Sokoloski.etal:13} together with optical spectroscopy suggesting bipolarity in its 2011 outburst \citep{Tofflemire.etal:13, Shore.etal:13}, radio observations of V959~Mon \citep{Chomiuk.etal:14,Linford.etal:15} revealing bipolar structure, 
optical, radio and X-ray imaging observations of asymmetry in the GK~Per nova remnant \citep{Bode.etal:04,Takei.etal:15}, and UV spectroscopic and IR interferometric evidence of a bipolar ejection in V339~Del \citep{Skopal.etal:14,Schaefer.etal:14}.

Consideration of the speed of flame propagation of a thermonuclear runaway on the white dwarf surface at the onset of a nova explosion by \citet{Fryxell.Woosley:82}
indicated a surface crossing time of about a day.  Recent multidimensional hydrodynamic simulations of the explosion initiation indicate that, even if ignition is point-like at the outset, flame propagation proceeds sufficiently rapidly that the explosion development is essentially spherically symmetric \citep{Casanova.etal:11}.  The shaping and collimation of nova explosions must then result from other processes.

\citet{Sokoloski.etal:08} favored an accretion-driven, jet-like scenario over shaping by circumstellar material as an explanation for highly collimated but transient features observed in mostly thermal radio emission after the 2006 RS~Oph blast.   We argue instead that 
bipolar structure is in general a result of the circumstellar environment.  This would be expected to be less collimated than the features seen by \citet{Sokoloski.etal:08}.  We speculate that the narrow jet-like appearance in that case might have been the result of shock-heating of slower-moving accretion-driven jet material. 

Detailed three-dimensional hydrodynamic modeling of the explosion of the recurrent nova U~Sco by \citet{Drake.Orlando:10} found that bipolar structure was effectively produced by the presence of an accretion disk that acts to focus the blast along the rotation axis of the binary.  In symbiotic systems, accretion might proceed with or without a disk, and the explosion can be additionally shaped by the wind of the companion.  The origin of the explosion is, of course, off-set from the origin of the wind by the binary separation.  The blast then traverses a range of density gradients at early times, from increasing to decreasing with the inverse square of the distance in directions  toward or away from  the evolved companion, respectively.    Since all nova progenitors either accrete through a disk or a dense wind, nova explosions should all be either asymmetric, collimated, or both. 

The geometry of the interaction of a nova with a dense companion wind was discussed by, e.g., \citet{Girard.Willson:87}, \citet{Drake.etal:09}, \citet{Nelson.etal:12} and \citet{Martin.Dubus:13}, and simulated in detail in three dimensions for the 2006 RS~Oph explosion by \citet{Walder.etal:08} and \citet{Orlando.etal:09}, and in two and three dimensions for the 2010 explosion of V407~Cyg by \citet{Orlando.Drake:12} and \citet{Pan.etal:15}, respectively.  
From line widths observed in near-infrared spectra, beginning 1.3 days after outburst discovery, \citet{Banerjee.etal:14} found the blast wave to be already evolving through a Sedov-Taylor deceleration  phase.  
The Sedov relations for  blast wave evolution noted in Section~\ref{s:xcomp} above 
imply that the shock proceeds much more rapidly away from the companion star ($w=2$) than toward it ($w=-2$).  Until the shock overtakes the evolved companion, its resulting shape in cross-section can resemble a cardioid whose cusp is aligned along the axis joining the explosion origin on the white dwarf and the evolved companion.  The much faster expansion in the direction away from the evolved companion produces a one-sided lobe which under certain circumstances might be considered jet-like. 

\citet{Orlando.etal:09} confirmed that the 2006 RS~Oph blast required an additional orbital plane density enhancement that was deduced from models by \citet{Walder.etal:08}.  Similar deductions were made for the V407~Cyg outburst (\citealt{Orlando.Drake:12,Martin.Dubus:13,Pan.etal:15}).  It seems likely that the V745 Sco blast would have been characterized by a similar density enhancement and we would therefore expect a bipolar structure for the forward-moving shock.  Predicting the morphology of the dominant X-ray emitting plasma is less straightforward: contributions can come from shocked circumstellar medium, the shocked equatorial density enhancement, and ejecta heated in the reverse shock.  \citet{Orlando.etal:09} found X-ray emission from their best-fit RS~Oph explosion model to be roughly equally divided between shocked circumstellar medium and shocked ejecta on day 13.9, with the equatorial density enhancement playing a large role.  In the case of V407~Cyg, X-ray emission in the \citet{Orlando.Drake:12} model was dominated by circumstellar medium plasma {\it behind} the red giant that was heated by the converging shock focussed to some extent by the wind density gradient.  Again, an equatorial density enhancement was required to fully understand the observed blast wave behavior.  

Our three-dimensional schematic models matching the observed line profiles suggested two possible configurations: an equatorial ring-like structure, or a one-sided polar cap.  The former requires a low-inclination and the latter a high inclination system.  In the absence of more detailed information on the system orbital parameters, further inference is not possible.  One clue as to the binary orbit configuration might lie in the difficulty in ascertaining the period from a considerable body of photometric data \citep{Schaefer:10,Mroz.etal:14}.  As noted by \citet{Mroz.etal:14}, this would favor a low inclination with the orbital plane close to that of the sky for which line-of-sight orbital photometric modulation would be small, and our ``ring-like" configuration.  Such an emission pattern might be produced by interaction of the blast with an equatorial density enhancement.  Further speculation requires exploration using detailed hydrodynamic simulations.

The shock wave evolution will not only depend on the gas distribution, but also on the state of the gas into which the shock system is running.  Some initial flash ionization of the circumstellar medium is likely to have occurred at the onset of the outburst, combined with further ionization from the SSS phase.  The degree of ionization is likely to follow the density distribution to some extent, with the dense, equatorial material being self-shielded and less ionized than lower density material at higher latitudes. 
Such effects may provide a resolution to the problem that a single one-dimensional model cannot account for both the 
observed temperatures and the widths of the H Pa$\beta$ line simultaneously. The $3\times 10^{43}$ erg
explosion provided the best match to the temperatures, but overestimated the H line width. A density
enhancement covering part of the blast wave surface would allow postshock recombination to proceed more
quickly, and would show up much brighter in H Pa$\beta$ emission than a shock running into uniform
density.  Additionally, if as has been argued previously, this equatorial density
enhancement resides principally in the plane of the sky, no Doppler shift between the forward
shocked H Pa $\beta$ and that from the reverse shock would be expected, in agreement with the
observations of \citet{Banerjee.etal:14}.

\subsection{Particle acceleration?}
\label{s:accel}

The deduction of \citet{Banerjee.etal:14} that the blast was already in the Sedov-Taylor phase only 1.3 days after discovery has important implications for the explosion parameters. 
The initial stage should be characterized by an episode of free spherical expansion until sufficient circumstellar material---a mass similar to that of the ejecta---has been swept up so as to begin to impede the blast.  At face value, the very short free expansion phase points to the ejected mass either being particularly small, or the circumstellar environment being particularly dense, or both.  One potential complication to this straightforward interpretation is the possibility of a significant energy sink in particle acceleration that \citet{Tatischeff.Hernanz:07} posed as an explanation for a short free expansion phase in the RS~Oph outburst.

The most egregious disagreement between observed and model parameters lies in the initial blast wave temperatures that are an order of magnitude higher than predicted during the first few days.  \citet{Tatischeff.Hernanz:07} have pointed out that particle acceleration through the first-order Fermi process can modify the shock behavior.  In particular, energy can be lost to accelerated particles that lowers the observed plasma temperature compared with that expected from the shock speed, while the post-shock compression ratio can be enhanced.   

The extremely high temperatures at early times might appear to be contrary to a significant particle acceleration energy sink, as does the agreement of the secular temperature trend with our blast wave models at later times.  Nevertheless, 
the $\gamma$-ray detection of the initial blast phase of V745~Sco \citep{Cheung.etal:15} indicates that significant particle acceleration did occur.  The $\gamma$-rays are thought to be produced either from the decay of neutral pions that arise in energetic proton collisions within the shock acceleration region, analogous to similar processes in supernovae, or from inverse Compton scattering and bremsstrahlung due to accelerated electrons \citep{Abdo.etal:10}.   The unexpected $\gamma$-rays found in non-symbiotic classical novae likely originate in the interaction of the blast with the immediate circumstellar environment, such as the accretion disk, with a period of a few days required from the explosion onset to engender sufficient particle acceleration to induce an observable $\gamma$-ray flux \citep{Ackermann.etal:14}. The $\gamma$-ray detection in V745~Sco coincided with the nova onset, suggesting that conditions in the immediate vicinity (in the stellar wind, equatorial density enhancement or an accretion disk) were conducive to rapid particle acceleration.  The extremely rapid drop in plasma temperature between days 4 and 10 might then be the result of energy lost to accelerated particle escape. 

\subsection{White dwarf mass}
\label{s:wdmass}

An estimate of the ejected mass based on the swept-up mass depends on the red giant mass loss rate, $\dot{M}_{wind}$, and wind speed, $v_{wind}$.  As noted in Section~\ref{s:anal_blast}, \citet{Banerjee.etal:14}  adopted $\dot{M}_{wind}=10^{-7}M_\odot$~yr$^{-1}$ and $v_{wind}=10$~km~s$^{-1}$, which, together with a shock moving at $4000$~km~s$^{-1}$ and traversing 3~AU in 1.3 days, implies a swept up mass at that time of $0.7\times 10^{-7}M_\odot$.  Since the accreted mass required to initiate a TNR, $\Delta  M$, depends on the ratio ${R_{WD}}^4/M_{WD}$ \citep{Fujimoto:82,Starrfield:89}, and $R_{WD}$ varies inversely with $M_{WD}$ to a power of greater than unity for $M_{WD}\geq 1M_\odot$, $\Delta M$ is a very sensitive function of the white dwarf mass.

Mass is lost both in the initial explosion, and in the aftermath until the end of the SSS phase as nuclear burning on the white dwarf surface at Eddington and super-Eddington luminosities drives a massive optically-thick wind.  The optically-thick wind models of \citet{Kato.Hachisu:94} for high-mass white dwarfs have mass loss rates of the order of $10^{20}$~g~s$^{-1}$ during the SSS phase.  For V745~Sco, with a SSS phase lasting about 10 days, the total mass loss is $\dot{M}_{wind}\sim 5\times 10^{-8}M_\odot$. For a very brief interval prior to SSS onset, the mass loss rate can  be higher still, such that it is not unreasonable to suppose that the mass lost in the initial explosion and subsequent wind are similar, $M_{eject}\sim M_{wind}$. 

\begin{figure*}
\begin{center}
\includegraphics[angle=0,width=6.2in]{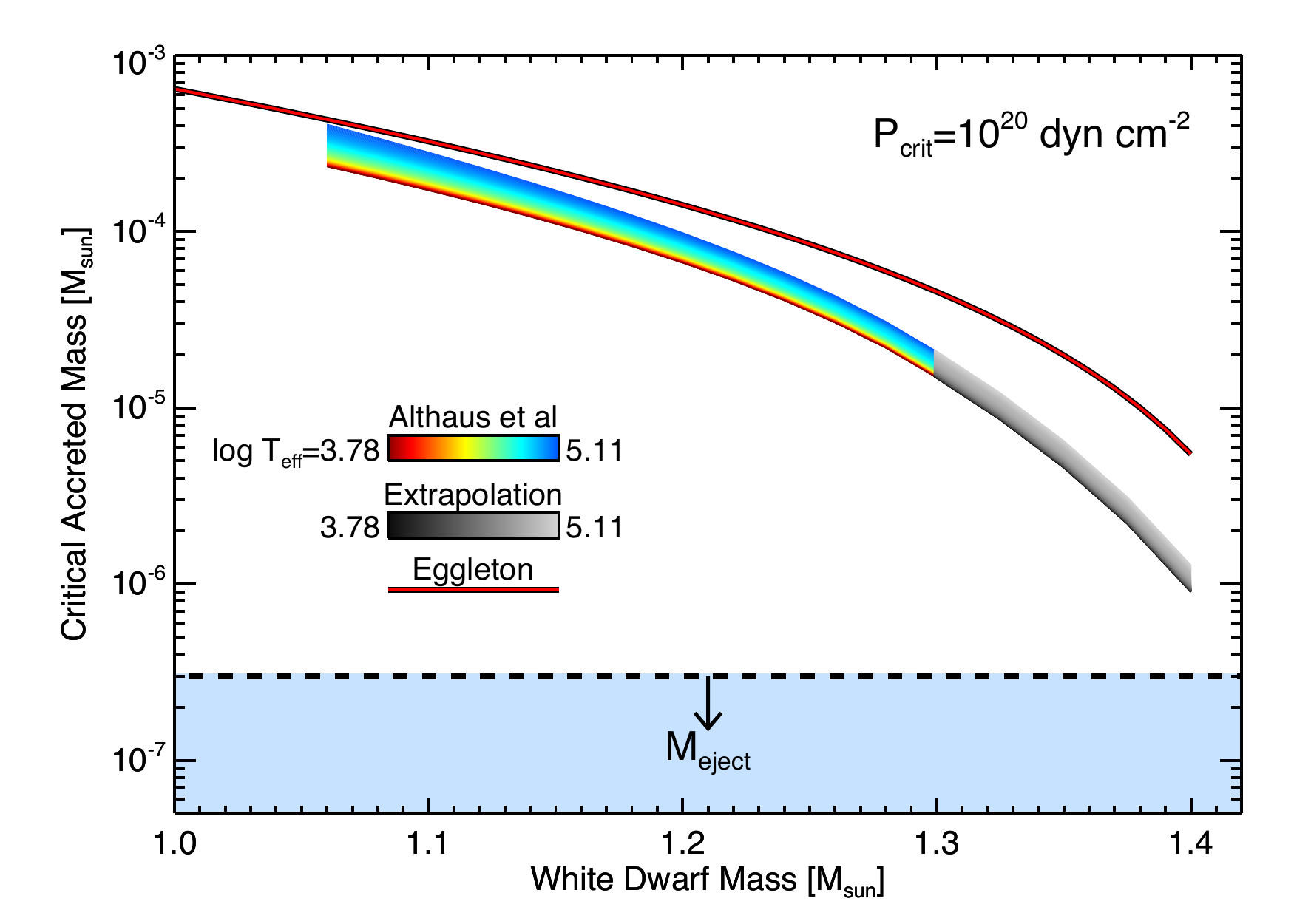}
\end{center}
\caption{The accreted mass required for TNR as a function of white dwarf mass, assuming a critical pressure of $P_{crit}=10^{20}$~dyn~cm$^{-2}$ and the mass-radius relations of Eggleton, as reported in \citet{Truran.Livio:86}, and  \citet{Althaus.etal:05}.  The dependence of the latter on stellar effective temperature is also illustrated, while the greyscale segment shows its linear extrapolation to higher masses.  The upper limit to the total ejected mass adopted here is indicated by the lower shaded region.
\label{f:mcrit}}
\end{figure*}

We compare the mass estimated to have been lost in the V745~Sco outburst with the mass required to initiate TNR assuming a critical pressure of $10^{20}$~dyn \citep{Starrfield:89} in Figure~\ref{f:mcrit}.  We adopt an upper limit of $M_{eject}+ M_{wind}\leq 3\times 10^{-7}M_\odot$, and the white dwarf mass-radius relation of \citet{Althaus.etal:05}, together with that attributed to Eggleton as used by, e.g., \citet{Truran.Livio:86}.   The required accreted mass for TNR, even for white dwarf masses approaching $1.4M_\odot$, is about an order of magnitude higher than our estimates.  This points both to the white dwarf being very close to the Chandrasekhar limit and to an inevitable net gain of mass of the order of a few $\times 10^{-7}M_\odot$ from each nova cycle.   A similar conclusion regarding both the stellar mass and mass gain was deduced by \citet{Kato:99} based on light curve modeling of the 1989 outburst.  \citet{Banerjee.etal:14} also point to the very high initial ejection velocity as supporting evidence that the ejected mass was relatively small and the white dwarf massive, while \citet{Page.etal:15} note that the remarkably short SSS phase also points to a very high white dwarf mass.

It is tempting to use the abundance results from the model parameter estimation described in Section~\ref{s:abuns} to place constraints on the composition of the underlying white dwarf.  Nova ejecta can be significantly enhanced in white dwarf material as a result of convective dredge-up during the thermonuclear runway.  This is betrayed by the presence of large enhancements in the spectra of classical nova explosion remnants in the ``nebular phase'' of C, N and O in the case of CO white dwarfs, and also Ne in the case of ONeMg white dwarfs \citep{Livio.Truran:94,Starrfield.etal:98}.  \citet{Mason:11} points out that [Ne/O]$>0$ for ONeMg novae, implying Ne enhancements relative to O by as much as fifty times the solar value.
At face value, the lack of obvious Ne enhancement in the V745~Sco {\it Chandra} spectra points to a CO white dwarf.  
However, unlike the case of classical novae with unevolved companions, the spectra of symbiotic novae like V745~Sco are complicated by the presence of the red giant wind.  The spectra analyzed here are of the early-time shocked wind system in which the ejecta may or may not contribute significant emission.
Without more detailed understanding of the blast wave and its evolution it is not possible to determine the relative contributions of the shocked circumstellar medium and shocked ejecta to the X-ray spectrum.  Simulations of the RS~Oph blast by \citet{Orlando.etal:09} indicated that ejecta could have contributed up to half of the observed line emission.  Further inference for V745~Sco must await similar detailed simulations.  In the absence of strong spectral features of Ne that would signal an ONeMg white dwarf, V745~Sco would be a Type 1a supernova (SN1a) progenitor candidate.

\section{CONCLUSIONS}

An analysis of {\it Chandra} HETG and LETG spectra of the recurrent symbiotic nova V745~Sco obtained between 16 and 17 days after outburst reveals a picture of a blast wave collimated by the circumstellar environment into which it expanded.  X-ray line profiles appear more peaked and triangular than the square-shaped profiles produced by spherical expansion.  Two schematic models match the observed shapes of the line profiles: expansion of a spherical cap at high inclination or of a ring-like structure at low inclination, such that motion is preferentially in a direction close to the plane of the sky. Two deviations from spherical symmetry in the circumstellar gas are likely to have played a role, as has been inferred for previous symbiotic novae: the off-set center of the blast within the $1/r^2$ density distribution of the red giant wind; and an equatorial density enhancement.

The X-ray lines are characterized by a FWZI~$\sim 2400$~km~s$^{-1}$, a FWHM$=1200\pm 30$~km~s$^{-1}$, and a net blueshift of $165 \pm 10$~km~s$^{-1}$.  Longer wavelength lines showed significant absorption by the central remnant in their red wings, with no photons detected in the O~VIII resonance line near 19~\AA\ red-ward of line center.   

At the time of the observations, the mean shocked plasma temperature was approximately 13 million degrees K, though plasma temperatures contributing significantly to the X-ray spectra ranged from 3-40~million degrees K.  Shocked plasma was cooling equally by radiative loss and adiabatic expansion, with a cooling time of the plasma at day 17 of approximately $10^6$~s, or 12 days. 

Comparison of an analytical blast wave model with the X-ray spectra and near-infrared H~I line widths of \citet{Banerjee.etal:14} suggests the explosion energy was of the order of 1--$3\times 10^{43}$~erg, and confirms the approximate ejected mass estimated by \citet{Banerjee.etal:14} of $10^{-7}M_\odot$.  Confronted with the accreted mass required for TNR being an order of magnitude larger than the mass lost, we confirm earlier suggestions \citep[e.g.][]{Kato:99} that V745~Sco is gaining several $10^{-7}M_\odot$ in each nova cycle and is a likely SN1a progenitor.

\acknowledgments

We thank the {\it Chandra X-ray Center} (CXC) Mission Planning team for rapid and timely scheduling of the ToO observations reported on here.  JJD and VK were funded by NASA contract NAS8-03060 to the CXC and thank the director, B.~Wilkes, and the CXC science team for advice and support.  LD and MH acknowledge the support of the Spanish Ministry of Economy and Competitivity (MINECO) under the grant ESP2014-56003-R. JML was supported by basic research
funds of the CNR.  KLP acknowledges funding from the UK
Space Agency. SS acknowledges partial support from NASA, NSF and {\it Chandra} grants to ASU. RDG was supported by NASA and the United States Air Force.  CEW acknowledges support from {\it Chandra} award G04-15023A. Finally, we also thank H.~Tananbaum for useful comments on the original manuscript.

\listofchanges

\end{document}